\documentclass[aps,prd,11pt,notitlepage,showpacs,nofootinbib,tightenlines]{revtex4-1}
\usepackage{amsmath}
\usepackage{amssymb}
\usepackage{epsfig}
\usepackage{graphicx}
\usepackage{bm}
\usepackage{times}
\usepackage{braket}
\usepackage{color}
\usepackage{slashed}
\usepackage{hyperref}
\definecolor{Red}{rgb}{1.,0.,0.}

\definecolor{Blue}{rgb}{0.,0.,1.}

\definecolor{nicered}{rgb}{0.7,0.1,0.1}
\definecolor{nicegreen}{rgb}{0.1,0.5,0.1}
\hypersetup{colorlinks,citecolor=nicegreen,linkcolor=nicered}
\makeatletter

\newcommand{\Rmnum}[1]{\expandafter\@slowromancap\romannumeral #1@}
\makeatother
\begin{document}
	%%%%%%%%%%%%%%%%%%%%%%%%%%%%%%%%%%%%%%%%%%%%%
	
	\newcommand{\beq}{\begin{eqnarray}}
	\newcommand{\eeq}{\end{eqnarray}}
	\newcommand{\ben}{\begin{enumerate}}
		\newcommand{\een}{\end{enumerate}}
	\newcommand{\non}{\nonumber\\ }
	
	\newcommand{\jpsi}{J/\Psi}
	
	\newcommand{\ppa}{\phi_\pi^{\rm A}}
	\newcommand{\ppp}{\phi_\pi^{\rm P}}
	\newcommand{\ppt}{\phi_\pi^{\rm T}}
	\newcommand{\ov}{ \overline }

	\newcommand{\zerot}{ {\textbf 0_{\rm T}} }
	\newcommand{\kt}{k_{\rm T} }
	\newcommand{\fb}{f_{\rm B} }
	\newcommand{\fk}{f_{\rm K} }
	\newcommand{\rk}{r_{\rm K} }
	\newcommand{\mb}{m_{\rm B} }
	\newcommand{\mw}{m_{\rm W} }
	\newcommand{\im}{{\rm Im} }
	
	\newcommand{\kks}{K^{(*)}}
	\newcommand{\acp}{{\cal A}_{\rm CP}}
	\newcommand{\pb}{\phi_{\rm B}}
	
	\newcommand{\xeba}{\bar{x}_2}
	\newcommand{\xsba}{\bar{x}_3}
	\newcommand{\peas}{\phi^A}
	
	\newcommand{\pvsl}{ p \hspace{-2.0truemm}/_{K^*} }
	\newcommand{\esl}{ \epsilon \hspace{-1.6truemm}/ }
	\newcommand{\psl}{ p \hspace{-1.6truemm}/ }
	\newcommand{\ksl}{ k \hspace{-1.6truemm}/ }
	\newcommand{\lsl}{ l \hspace{-1.6truemm}/ }
	\newcommand{\nsl}{ n \hspace{-1.6truemm}/ }
	\newcommand{\vsl}{ v \hspace{-1.6truemm}/ }
	\newcommand{\epsl}{\epsilon \hspace{-1.6truemm}/\,  }
	\newcommand{\bfkk}{{\bf k} }
	\newcommand{\calm}{ {\cal M} }
	\newcommand{\calh}{ {\cal H} }
	\newcommand{\calo}{ {\cal O} }
	
	%%---------------------------------------------------------
	%%%%%%%%%%%%%%%%%%%
	\def \appb{{\bf Acta. Phys. Polon. B }  }
	\def \cpc{ {\bf Chin. Phys. C } }
	\def \ctp{ {\bf Commun. Theor. Phys. } }
	\def \epjc{{\bf Eur. Phys. J. C} }
	\def \ijmpcs{{\bf Int. J. Mod. Phys. Conf. Ser.} }
	\def \jhep{{\bf J. High Energy Phys. } }
	\def \jpg{ {\bf J. Phys. G} }
	\def \mpla{{\bf Mod. Phys. Lett. A } }
	\def \npb{ {\bf Nucl. Phys. B} }
	\def \plb{ {\bf Phys. Lett. B} }
	\def \ppn{ {\bf Phys. Part. Nucl. } }
	\def \ppnp{{\bf Prog.Part. Nucl. Phys.  } }
	\def \pr{  {\bf Phys. Rep.} }
	\def \prc{ {\bf Phys. Rev. C }}
	\def \prd{ {\bf Phys. Rev. D} }
	\def \prl{ {\bf Phys. Rev. Lett.}  }
	\def \ptp{ {\bf Prog. Theor. Phys. }}
	\def \zpc{ {\bf Z. Phys. C}  }
	\def \jpg{ {\bf J.Phys. G}  }
	\def \ap{ {\bf Ann. of Phys}  }

%%%%%%%%%%%%%%%%%%%%%%%%%%%%%%%%%%%%%%%%%%%%%%%%%%%%
%%
\title{The next-to-leading order corrections to  $B \to \rho$ transition  in the $k_T$ factorization}
\author{Jun Hua$^{1}$} \email{546406604@qq.com}
\author{Ya-Lan Zhang$^{2}$} \email{zylyw@hyit.edu.cn}
\author{Zhen-Jun Xiao$^{1,3}$ } \email{xiaozhenjun@njnu.edu.cn}
	\affiliation{1.  Department of Physics and Institute of Theoretical Physics,
		Nanjing Normal University, Nanjing, Jiangsu 210023, People's Republic of China,}
	\affiliation{2. Department of Faculty of Mathematics and Physics,  Huaiyin Institute of Technology,
Huaian, Jiangsu 223001, People's Republic of China, }
	\affiliation{3. Jiangsu Key Laboratory for Numerical Simulation of Large Scale Complex Systems,
	Nanjing Normal University, Nanjing 210023, People's Republic of China}
	\date{\today}
	%\vspace{1cm}
%%===============================
\begin{abstract}
In this paper, we investigate the factorization hypothesis step by step for the exclusive process $B \to \rho$ at next-to-leading order (NLO)
with the collinear factorization approach, and then we extend our results to the $k_T$ factorization frame.
We show that the soft divergence from the specific NLO diagrams will cancel each other at the quark level,
while the remaining collinear divergence can be absorbed into the NLO wave functions.
The full NLO amplitudes can be factorized into  two parts: the NLO $B$ and $\rho$ meson wave functions containing
the collinear divergence and the leading order (LO) finite hard kernels.
We give the general form of the nonlocal hadron matrix for the NLO $B$ and $\rho$ meson wave functions
and all results of factorization for different twists' combinations.
\end{abstract}

%\vspace{1cm}

%\keywords{collinear factorization, $k_T$ factorization, next-to-leading-order correction, $B \to \rho$ transition form factor.}

\maketitle

\section{Introduction}

As the foundation of perturbative QCD formalism, the factorization theorem \cite{plb87-359,prd55-272,prp112-173} claims that hard part of the  QCD
interactions  is finite and can be calculated perturbatively, meanwhile the non-perturbative part can be factorized into the universal wave functions
defined in an infinite momentum frame. The perturbative QCD (PQCD) approach\cite{kt-fact} based on the $k_T$ factorization theorem
is proposed to eliminate endpoint divergence by picking up the transversal momentum of the propagators together with the Sudakov
resummation technique\cite{PQCD-fact1,prove}  .

In recent years, several exclusive processes $\pi\gamma^* \to \gamma(\pi)$, $\rho \gamma^* \to \pi$ and $B \to \gamma(\pi)l\bar{\nu}$ have
been investigated by many authors for example in Refs.~\cite{PQCD-fact1,prove,Zhang-Hua,PQCD-NLO-Li,PQCD-NLO-Cheng}
at the leading order (LO) and next-to-leading order (NLO). The calculations of the space like and time like form factors \cite{PQCD-NLO-timelike}
will be a great help for high precision studies  for B meson decays.

In this paper, taking the $\rho$ meson as one example for those light vector mesons,  we consider the $B \to \rho$ transition process,
give the proof of collinear factorization at the NLO level for leading twist parts.
Fierz identity and eikonal approximation will be taken into account to factorize the fermion flow and momentum flow respectively.
The power counting for various gamma matrices
we discussed in \cite{prd95-076005} are also considered in order to give the right color factors with triple gluon vertex diagrams.
By summing up the contributions from all sub-diagrams,
we can prove that the soft divergence will cancel each other between quark diagrams, and the remaining collinear divergence can also be
absorbed into the NLO meson wave functions.
The convolutions of the NLO wave function and hard kernel have two forms: the one has additional gluon emitted from
external quarks flowing into hard kernel; another has no additional gluon flowing into hard kernel.
We will finally write down the hadronic matrix elements for $B$ and $\rho$ meson wave functions with collinear parts in $k_T$ factorization.

This paper is organized as follows. In section \Rmnum{2}, we show the dynamical analysis and the leading order amplitudes
for $B \to \rho$ transition. In section \Rmnum{3}, we present the proof of collinear factorization for $B \to \rho$ transition
at the NLO step by step, and give the hadronic matrix elements for wave functions in $k_T$ factorization.
In section \Rmnum{4}, finally, a brief summary and some discussions will be given.

\section{Leading order hard kernel}

\begin{figure}[htbp]
\centering
\includegraphics[width=4in]{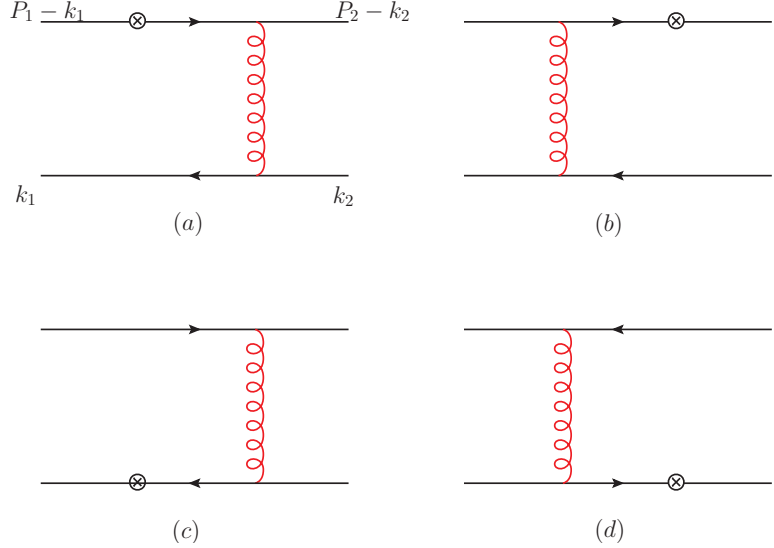}
\caption{The leading-order quark diagrams for the $B \to \rho$ transition form factors
with the symbol $\otimes$ representing the weak vertex. }
\label{fig:1}
\end{figure}

In this section, we calculate the leading order $O(\alpha_s)$ hard kernel for  $B \to \rho$  transition, and the topological diagrams of leading-order process
are displayed in Fig.~\ref{fig:1}. The definition of the kinematics in light cone coordinate are of the following form:
{\small
    \beq
	p_{1 \mu} &=& \frac{m_B}{\sqrt{2}}(1,1,\mathbf{0_T}), \qquad
	k_{1 \mu} = \frac{m_B}{\sqrt{2}}(x_1,0,\mathbf{0_T});\non
	\epsilon_{1 \mu}(L) &=& \frac{1}{\sqrt{2}}(1,0,\mathbf{0_{T}}),~~~~~
	\quad\epsilon_{1 \mu}(T)  =  (0,0,\mathbf{1_{T}});
	\label{eq:kinematics-1}
	\eeq
	\beq
	p_{2 \mu} &=& \frac{\eta m_B}{\sqrt{2}}(0,1,\mathbf{0_T}),~~~~~
	\quad k_{2 \mu}  =  \frac{\eta m_B}{\sqrt{2}}(0,x_2,\mathbf{0_T});\non
	\epsilon_{2 \mu}(L) &=& \frac{1}{\sqrt{2}\gamma_{\rho}}(-\gamma_{\rho}^2,1,\mathbf{0_{T}}),~~~~~
    \epsilon_{2 \mu}(T)  =  (0,0,\mathbf{1_{T}}),
	\label{eq:kinematics-2}
    \eeq}
where the $m_B$ is the mass of $B$-meson, the momentum $P_1$ is chosen as ${m_B}/\sqrt{2} (1,1,\mathbf{0_T})$, both $p^+$ and $p^-$
parts count due to the large scale of $m_B$. The energy fraction $\eta$ of the $\rho$ meson in large recoil region is of order $\mathbf{1}$, and the
polarization vector with the definition $\gamma_{\rho} = m_{\rho}/Q $ is defined by condition $\epsilon^2_i(L/T)= -1$. The momentum transfer
squared $Q^2=-(p_1-p_2)^2 \quad (Q^2>0)$ is taken to represent evolution behaviors of form factors.

The leading twist (twist-2) wave functions of B and $\rho$ mesons are chosen in the same form as those in Refs.~\cite{WfB,WFsrho}:
{\small
	\beq
	\Phi_{B}(x_1,p_1) &=& \frac{1}{\sqrt{2N_c}}(\psl_1 + M_B) \left[\frac{\nsl_+}{\sqrt{2}}\phi^+_B(x_1) + \frac{1}{\sqrt{2}}\left
	(\nsl_- - k^+_1\gamma^{\nu}_{\perp} \frac{\partial}{\partial \mathbf{k}^{\nu}_{1T}}\right)\phi^-_B(x_1) \right], \non
	\label{eq:wfB}
	\Phi_{\rho}(p_2,\epsilon_{2L})&=&\frac{i}{\sqrt{2N_{c}}}\left [ M_{\rho} \esl_{2L} \phi_{\rho}(x_{2})\right ],\non
%	+\esl_{2L} \psl_{2}\phi^{t}_{\rho}(x_{2}) + M_{\rho} \phi^{s}_{\rho}(x_{2})\right ],\non
	\Phi_{\rho}(p_2,\epsilon_{2T})&=&\frac{i}{\sqrt{2N_{c}}}
	\left [M_{\rho} \esl_{2T} \phi^{v}_{\rho}(x_{2})\right ],
%	+\esl_{2T} \psl_{2}\phi^{T}_{\rho}(x_{2})
%	+M_{\rho} i \epsilon_{\mu' \nu \rho \sigma}  \gamma_{5}\gamma^{\mu'} \epsilon^{\nu}_{2T} v^{\rho} n^{\sigma} \phi^{a}_{\rho}(x_{2})\right ],
\label{eq:wfrho}
\eeq}
where $N_c=3$ is the number of colors,  $n_{+}=(1,0,0)$ and $n_{-}= (0,1,0)$ are the unit vector  in the light cone coordinate.
The wave functions of the vector $\rho$ meson contain both the longitudinal and transversal component  $\Phi_{\rho}(p_2,\epsilon_{2L})$ and
$\Phi_{\rho}(p_2,\epsilon_{2T})$. To insure the gauge invariance and the accuracy
of calculation, both contributions from $\phi^+_B$ and $\phi^-_B$ will be taken into account in this work.

Because of the symmetry relations, the contributions from Fig.~\ref{fig:1}(c,d) can be simply derived from those for Fig.~\ref{fig:1}(a,b)
by exchange of momentum fraction for up (down) quarks. Then the LO transition amplitude of Fig.~\ref{fig:1}(a) and (b) can be simplified to the following forms:
{\small
    \beq
    H^{(0)}_{aL} &=& -\frac{ieg^2C_F}{2} \mathbf{Tr}  \bigg{\{} \frac{\gamma^{\alpha}[M_{\rho} \esl_{2L} \phi_{\rho}(x_{2})] \gamma_{\alpha} (\psl_2 - \ksl_1) \gamma_{\mu}
    (\psl_1 + m_B) \gamma_{5}\left[\frac{\nsl_+}{\sqrt{2}}\phi^+_B(x_1) + \frac{\nsl_-}{\sqrt{2}}\phi^-_B(x_1)\right]}
    {(p_2-k_1)^2(k_1-k_2)^2} \bigg{\}} \non
    &=&-ieg^2C_Fm_B\mathbf{Tr} \left[\frac{\esl_{2L} \ksl_1 \gamma_{\mu} \gamma_{5} \frac{\nsl_+}{\sqrt{2}} \phi^+_B(x_1) }{(p_2-k_1)^2(k_1-k_2)^2}  \right],
    \label{eq:Hal}
    \eeq}
{\small
   \beq
   H^{(0)}_{bL} &=&-ieg^2C_Fm_B\mathbf{Tr} \left[ \frac{ M_{\rho} \esl_{2L} \phi_{\rho}(x_{2}) \gamma_{\mu} \ksl_2 \frac{\nsl+}{\sqrt{2}}\phi^+_B(x_1)\gamma_{5}
   + M_{\rho} \esl_{2L} \phi_{\rho}(x_{2}) \gamma_{\mu} m_b \phi^-_B(x_1) \gamma_{5}}
   {[(p_1-k_2)^2-m^2_b](k_1-k_2)^2}\right],     \label{eq:Hbl} \\
   H^{(0)}_{bT} &=&-ieg^2C_Fm_B\mathbf{Tr} \left[\frac{\Big[\esl_{2T} \psl_{2}\phi^{T}_{\rho}(x_{2})\Big]\gamma_{\mu}\psl_1
   \Big[\phi^+_B(x_1)+\phi^-_B(x_1)\Big]\gamma_{5}
   - \Big[\esl_{2T} \psl_{2}\phi^{T}_{\rho}(x_{2})\Big]\gamma_{\mu}m_b \frac{\nsl_+}{\sqrt{2}}\phi^+_B(x_1)\gamma_{5}}
   {\Big[(p_1-k_2)^2-m^2_b\Big](k_1-k_2)^2} \right],	
   \label{eq:Hbt}
\eeq}
where  $C_F=4/3$ is the color factor. Since  $H^{(0)}_{aT}=0$ for Fig.~\ref{fig:1}(a), we do not show $H^{(0)}_{aT}$ explicitly in Eq.~(\ref{eq:Hal}).
For $H^{(0)}_{aL}$, only the $\ksl_1$ in quark propagator and $m_B$ component in B wave function have contributions, which also requires $\gamma^{\alpha}$ and
$\gamma_{\mu}$ should be $\gamma_{\perp}$ and $\gamma^+$.
$H^{(0)}_{bL}$ contains two components corresponding to the two parts of the $B$ wave
function $\phi^+_B$ and $\phi^-_B$ respectively, which leads to different restrictions on the  vertex with $\gamma_{\alpha} $ and $\gamma_{\mu}$.
Similarly, $H^{(0)}_{bT}$ contains three components, two for $\phi^+_B$ and one for $\phi^-_B$.
Every part of the combinations requires unique choice for $P_1$ momentum $(P^-_1/P^+_1)$ and gamma matrix
$\gamma_{\mu}$ and $\gamma_{\alpha}$. We should consider each part independently.

\section{Factorization Of $B \to \rho$ At Next-To-Leading Order}

In this section, we will give the proof of collinear factorization for the exclusive process $B \to \rho$ at NLO level, considering both momentum space and color
space factorization but neglecting the transverse momentum first. At the end of this section, we will pick up again the transverse momentum and discuss
how to deal with the infrared contribution carrying by $k_T$-related part.

The fermion flow in momentum space can be factorized by inserting the Fierz identity
\beq
	I_{ij}I_{lk}   &=& \frac{1}{4}I_{ik}I_{lj} + \frac{1}{4}(\gamma_{5})_{ik}(\gamma_{5})_{lj}
	+ \frac{1}{4}(\gamma^{\alpha})_{ik}(\gamma^{\alpha})_{lj}\non
	&&              + \frac{1}{4}(\gamma_{5}\gamma^{\alpha})_{ik}(\gamma_{\alpha}\gamma_{5})_{lj}
	+ \frac{1}{8}(\sigma^{\alpha\beta})_{ik}(\sigma_{\alpha\beta})_{lj},	\label{eq:Fierz}
\eeq
into the proper place of the matrix elements,  here $\sigma^{\alpha\beta} = i [\gamma^{\alpha},\gamma^{\beta}]/2$.
The different terms on the right hand side correspond to the contributions from  different twists of $B (\rho)$ meson.
The eikonal approximation is taken before inserting the Fierz identity to reformulate the singularity propagator
into a simplified form and reduce the gamma matrix for convenience meanwhile.
All the possible diagrams for a gluon radiation should be considered
and finally be resumed to collect the color factors and maintain the gauge invariance.

\subsection{Collinear factorization for the NLO(${\calo}(\alpha^2_s)$) corrections to $H^{(0)}_{aL}$}

\begin{figure}[htbp]
	\centering
	\includegraphics[width=4.5in]{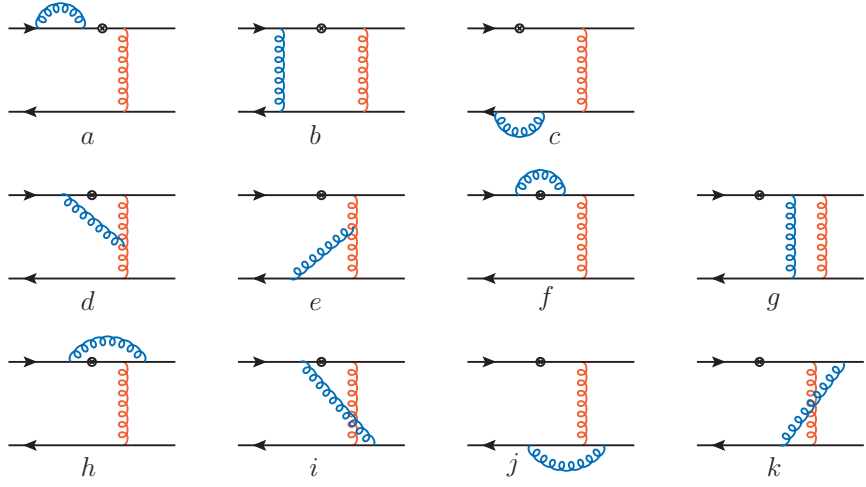}
	\caption{The quark diagrams for NLO corrections to Fig.~\ref{fig:1}(a) with additional gluon emitted from the quark(antiquark) lines of initial B meson }
	\label{fig:2}
\end{figure}

We will only show the factorization of the NLO corrections for Fig.~\ref{fig:1}(a,b), for the reason that Fig.~\ref{fig:1}(c,d) can be derived from
Fig.~\ref{fig:1}(a,b) by symmetry. Taking into account the different twists' combinations, there exist six different structures from the decay amplitudes as listed
Eqs.~(\ref{eq:Hal}),(\ref{eq:Hbl}). Each structure has two terms corresponding to the additional gluon lines emitted from the initial $B$ meson or the
finial $\rho$ meson.

The topological diagrams for NLO corrections to Fig.~\ref{fig:1}(a) with additional gluon emitted from  initial B meson are showed in Fig.~\ref{fig:2}.
Generally, there exists two kinds of infrared divergence: (a) the soft divergence when the momentum of additional gluon is small at $l = (l^+,l^-,l_{\perp}) \sim
(\lambda,\lambda,\lambda)$ for $\lambda \sim \Lambda_{QCD}$; and (b) the collinear divergence when the additional gluon is parallel with one longitudinal
direction $l = (l^+,l^-,l_{\perp}) \sim (\lambda^2 / Q^2,Q,\lambda^2)$\cite{jhep0802-002,npb844-199}. 
The third kind infrared divergence appeared in the Glauber region of NLO spectator amplitudes of B meson 
two-body nonleptonic decays\cite{prd90-074018,prd91-114019} does not exist here. We rewrite the Eq.~(\ref{eq:Hal})
explicitly in the following form:
{\small
	\beq
	H^{(0)}_{aL} &=& -\frac{ieg^2C_F}{2} \mathbf{Tr}
	\bigg{\{} \frac{\gamma^{\alpha}[M_{\rho} \esl_{2L} \phi_{\rho}(x_{2})] \gamma_{\alpha} ( - \ksl_1) \gamma_{\mu} ( m_B)
\gamma_{5}\left[\frac{\nsl_+}{\sqrt{2}}\phi^+_B(x_1) \right]}
	{(p_2-k_1)^2(k_1-k_2)^2} \bigg{\}},
	\label{eq:GaL0}
	\eeq}
in which we can see only $\ksl_1$ in internal quark propagator and $m_B$ part in $B$ meson wave function can contribute in this channel,
and $\gamma_{\mu}$ and $\gamma^{\alpha}$ should be $\gamma^+$ and $\gamma_{\perp}$.

We can straightly write down the NLO amplitudes of Fig.~\ref{fig:2}(a,b,c)  since  no radiated gluon lines attached to internal line,
and therefore no momentum will flow into the structure of LO hard kernel. So factorization can be derived simply by inserting the Fierz identity,
neatly cutting on the external quark (antiquark) line.
{\small
    \beq
    G^{(1)}_{aL,2a} &=& \frac{1}{2} \frac{eg^4C^2_F}{2} \mathbf{Tr}  \bigg{\{}
    \frac{\gamma^{\alpha} \left[M_{\rho} \esl_{2L} \phi_{\rho}(x_{2})\right] \gamma_{\alpha} ( - \ksl_1) \gamma_{\mu}
    (\psl_1-\ksl_1+m_b) \gamma^{\nu} (\psl_1-\ksl_1-\lsl+m_b) \gamma_{\nu} ( m_B) \gamma_{5}\left[\frac{\nsl_+}{\sqrt{2}}\phi^+_B(x_1) \right]  }
    {(p_2-k_1)^2(k_1-k_2)^2[(p_1-k_1)^2-m^2_b][(p_1-k_1-l)^2-m^2_b] l^2}  \bigg{\}}  \non
    &=&\frac{1}{2} \phi^{(1)}_{B,a} \otimes G^{(0)}_{aL}(x_1,x_2), 	\label{eq:GaL-2a} \\
    %%%%%
    G^{(1)}_{aL,2b} &=& -\frac{eg^4C^2_F}{2} \mathbf{Tr}  \bigg{\{}
    \frac{ \gamma^{\nu} (\ksl_1-\lsl) \gamma^{\alpha} \left[M_{\rho} \esl_{2L} \phi_{\rho}(x_{2})\right] \gamma_{\alpha}
    (-\ksl_1+\lsl) \gamma_{\mu} (\psl_1-\ksl_1+\lsl+m_b) \gamma_{\nu} ( m_B) \gamma_{5}\left[\frac{\nsl_+}{\sqrt{2}}\phi^+_B(x_1) \right]  }
    {(p_2-k_1+l)^2(k_1-k_2-l)^2(k_1-l)^2[(p_1-k_1+l)^2-m^2_b] l^2}  \bigg{\}}  \non
    &=& \phi^{(1)}_{B,b} \otimes G^{(0)}_{aL}(\xi_1,x_2), 	\label{eq:GaL-2b}\\
    %%%%%
    G^{(1)}_{aL,2c} &=& \frac{1}{2}\frac{eg^4C^2_F}{2} \mathbf{Tr}  \bigg{\{}
    \frac{ \gamma^{\nu} (\ksl_1-\lsl) \gamma_{\nu} (\ksl_1) \gamma^{\alpha} \left[M_{\rho} \esl_{2L} \phi_{\rho}(x_{2})\right]
    \gamma_{\alpha} (-\ksl_1) \gamma_{\mu} (m_B) \gamma_{5}\left[\frac{\nsl_+}{\sqrt{2}}\phi^+_B(x_1) \right]  }
    {(p_2-k_1)^2(k_1-k_2)^2(k_1-l)^2(k_1) l^2}  \bigg{\}}  \non
    &=& \frac{1}{2} \phi^{(1)}_{B,c} \otimes G^{(0)}_{aL}(x_1,x_2),     \label{eq:GaL-2c}
    \eeq}
with
{\small
    \beq
    \phi^{(1)}_{B,a} &=& \frac{-ig^2C_F}{4} \frac{\gamma_{5} \gamma^{\rho} \gamma_{\rho} \gamma_{5} (\psl_1-\ksl_1+m_b)
    \gamma^{\nu} (\psl_1-\ksl_1-\lsl+m_b) \gamma_{\nu} }{[(p_1-k_1)^2-m^2_b][(p_1-k_1-l)^2-m^2_b] l^2},  	\label{eq:phi1-ba}\\
    %%%%
    \phi^{(1)}_{B,b} &=& \frac{ig^2C_F}{4} \frac{\gamma_{5} \gamma^{\rho} (\psl_1-\ksl_1+\lsl+m_b) \gamma^{\nu} \gamma_{5}
    \gamma_{\rho} \gamma_{\nu} (\ksl_1-\lsl) }{(k_1-l)^2[(p_1-k_1+l)^2-m^2_b] l^2} ,	\label{eq:phi1-bb}\\
    %%%%%
    \phi^{(1)}_{B,c} &=& \frac{-ig^2C_F}{4} \frac{ \gamma_{5} \gamma_{\rho} \gamma^{\nu} (\ksl_1-\lsl) \gamma_{\nu} (\ksl_1) \gamma^{\rho} \gamma_{5} }
    {(k_1-l)^2 (k_1) l^2}.
    \label{eq:phi1-bc}
    \eeq}
The factor $1/2$ appeared in Eqs.~(\ref{eq:GaL-2a},\ref{eq:GaL-2c}) comes from the symmetry of Fig.~\ref{fig:2}(a) and \ref{fig:2}(c).
From the QCD diagrams and dynamics, no independent soft divergences
exist in reducible diagrams, then the collinear divergences can be absorbed into NLO wave functions concisely.
These infrared singularities from
amplitudes of Figs.~\ref{fig:2}(a,b,c) are absorbed into NLO wave functions $\phi^{(1)}_{B,i} (i=a,b,c)$ and can be re-expressed as in Eqs.~(\ref{eq:phi1-ba}-\ref{eq:phi1-bc}).
For the reason that the infrared divergence in reducible diagrams without mixing can be simply matched one by one between QCD diagrams and effective
diagrams\cite{PQCD-NLO-Li}, these results are independent out of  these irreducible diagrams. So, we just pay more attention to irreducible diagrams.

For Figs.~\ref{fig:2}(d,e) we have
{\small
 \beq
G^{(1)}_{aL,2d} &=& \frac{ieg^4Tr[T^aT^bT^c]f_{abc}}{2N_c}   \non
&& \times \mathbf{Tr} \bigg{\{} \frac{\gamma^{\alpha}\left[M_{\rho} \esl_{2L} \phi_{\rho}(x_{2})\right] \gamma^{\beta} (\psl_{2}-\ksl_1+\lsl)
\gamma_{\mu} (\psl_1-\ksl_1+\lsl+m_b) \gamma^{\gamma}  (m_B) \gamma_{5}\left[\frac{\nsl_+}{\sqrt{2}}\phi^+_B(x_1)\right] F_{\alpha \beta \gamma} }
    {(k_1-k_2)^2(k_1-k_2-l)^2(p_2-k_1+l)^2[(p_1-k_1+l)^2-m^2_b]l^2}  \bigg{\}}   \non
    &=& \frac{9}{8} \phi^{(1)}_{B,d} \otimes \left[G^{(0)}_{aL}(\xi_1,x_1,x_2)-G^{(0)}_{aL}(\xi_1,x_2) \right], \non
    %%%%%
 G^{(1)}_{aL,2e} &=& \frac{ieg^4Tr[T^aT^bT^c]f_{abc}}{2N_c}   \non
    && \times \mathbf{Tr} \bigg{\{} \frac{\gamma^{\alpha}(\ksl_1-\lsl) \gamma^{\beta} \left[M_{\rho} \esl_{2L} \phi_{\rho}(x_{2})\right]
    \gamma^{\gamma} (\psl_{2}-\ksl_1) \gamma_{\mu} (m_B)\gamma_{5}\left[\frac{\nsl_+}{\sqrt{2}}\phi^+_B(x_1)\right] F_{\alpha \beta \gamma} }
    {(k_1-k_2)^2(k_1-k_2-l)^2(p_2-k_1)^2(k_1-l)^2l^2}  \bigg{\}}   \non
    &=& \frac{9}{8} \phi^{(1)}_{B,e} \otimes \left[G^{(0)}_{aL}(x_1,x_2)-G^{(0)}_{aL}(x_1,\xi_1,x_2) \right],
    \label{eq:GaLde}
    \eeq}
with
{\small
\beq
	\phi^{(1)}_{B,d} &=& \frac{-ig^2C_F}{4} \frac{ \gamma^+\gamma_{5}  (\psl_1-\ksl_1+\lsl+m_b) \gamma^{\rho} \gamma_{5}\gamma^- }{[(p_1-k_1+l)^2
-m^2_b] l^2} \frac{\nu_{\rho}}{\nu \cdot l},  \non
	%%%%
	\phi^{(1)}_{B,e} &=& \frac{ig^2C_F}{4} \frac{ \gamma_{5}\gamma^- \gamma^{\rho} (\ksl_1-\lsl)  \gamma^+\gamma_{5} }{(k_1-l)^2 l^2}
\frac{\nu_{\rho}}{\nu \cdot l}.
	\label{eq:WFaLde}
	\eeq}
For Fig.~\ref{fig:2}(d) where the triple gluon vertex appeared, as we examined  in Ref.~\cite{prd95-076005}, only the terms in
$G^{(1)}_{aL,2d}$ proportional to $g_{\alpha \beta}$ in $F_{\alpha \beta \gamma}
= g_{\alpha \beta} (2k_2-2k_1+l)_{\gamma} + g_{\beta \gamma} (k_1-k_2-2l)_{\alpha}+g_{\gamma \alpha} (k_1-k_2+l)_{\beta}$
give the right NLO corrections to the LO hard kernel. The other terms are power suppressed by power counting of matrix elements and also
unphysical in topological diagrams.
For Fig.~\ref{fig:2}(e), similarly, only the terms  proportional to $g_{\beta \gamma}$ in $F_{\alpha \beta \gamma} = g_{\alpha \beta}
(k_1-k_2-2l)_{\gamma} + g_{\beta \gamma} (2k_2-2k_1+l)_{\alpha}+g_{\gamma \alpha} (k_1-k_2+l)_{\beta}$ contribute effectively.
Then the NLO $B$ meson wave function containing infrared divergence with $``l"$ emitted from quark (antiquark) line can be represented as in Eq.~(\ref{eq:WFaLde}).
Because of the large scale of $m_b$, the $\phi^{(1)}_{B,d}$ contains the $m_b$ term in propagator when $``l"$ emits from the $b$ quark .

The factorization of amplitudes for remaining diagrams   Fig.~\ref{fig:2}(f-k) can be expressed as follows:
{\footnotesize
	\beq
    G^{(1)}_{aL,2f} &=& - \frac{eg^4Tr[T^aT^aT^cT^c]}{2N_c}  \non
    && \times \mathbf{Tr} \bigg{\{} \frac{\gamma^{\alpha} \left[M_{\rho} \esl_{2L} \phi_{\rho}(x_{2})\right] \gamma_{\alpha} (\psl_2-\ksl_1) \gamma^{\nu}
    (\psl_2-\ksl_1 + \lsl) \gamma_{\mu} (\psl_1-\ksl_1+\lsl+m_b) \gamma_{\nu}  (m_B) \gamma_{5}\left[\frac{\nsl_+}{\sqrt{2}}\phi^+_B(x_1) \right] }
    {(p_2-k_1)^2(k_1-k_2)^2(p_2-k_1+l)^2[(p_1-k_1+l)^2-m^2_b] l^2}  \bigg{\}} \non
    && \sim 0 ,   \label{eq:GaLf-f} \\
    %%%%%%
    G^{(1)}_{aL,2g} &=& \frac{eg^4Tr[T^aT^aT^cT^c]}{2N_c}  \non
&& \times \mathbf{Tr} \bigg{\{} \frac{\gamma^{\nu} (\ksl_1 - \lsl) \gamma^{\alpha} \left[M_{\rho} \esl_{2L} \phi_{\rho}(x_{2})\right]
\gamma_{\alpha} (\psl_2-\ksl_1+\lsl)    \gamma_{\nu}(\psl_2-\ksl_1)  \gamma_{\mu} (m_B)\gamma_{5}\left[\frac{\nsl_+}{\sqrt{2}}\phi^+_B(x_1) \right] }
    {(p_2-k_1)^2(k_1-k_2-l)^2(p_2-k_1+l)^2(k_1-l)^2 l^2}  \bigg{\}} \non
    && \sim 0,  \label{eq:GaLf-g}\\
    %%%%%%
    G^{(1)}_{aL,2h} &=& - \frac{eg^4Tr[T^aT^cT^aT^c]}{2N_c}  \non
    && \times \mathbf{Tr} \bigg{\{} \frac{\gamma^{\alpha} \left[M_{\rho} \esl_{2L} \phi_{\rho}(x_{2})\right] \gamma^{\nu}  (\psl_2-\ksl_2+\lsl)
    \gamma_{\alpha} (\psl_2-\ksl_1 + \lsl) \gamma_{\mu} (\psl_1-\ksl_1+\lsl+m_b) \gamma_{\nu} (m_B) \gamma_{5}\left[\frac{\nsl_+}{\sqrt{2}}\phi^+_B(x_1) \right] }
    {(p_2-k_1+l)^2(k_1-k_2)^2(p_2-k_2+l)^2[(p_1-k_1+l)^2-m^2_b] l^2}  \bigg{\}} \non
    &=&  - \frac{1}{8} \phi^{(1)}_{B,d} \otimes G^{(0)}_{aL}(\xi_1,x_1,x_2),   \label{eq:GaLf-h}\\
    %%%%%%
    G^{(1)}_{aL,2i} &=&  \frac{eg^4Tr[T^aT^cT^aT^c]}{2N_c}  \non
    && \times \mathbf{Tr} \bigg{\{} \frac{\gamma^{\alpha} (\ksl_2+\lsl) \gamma^{\nu} \left[M_{\rho} \esl_{2L} \phi_{\rho}(x_{2})\right]
    \gamma_{\alpha} (\psl_2-\ksl_1 + \lsl) \gamma_{\mu} (\psl_1-\ksl_1+\lsl+m_b) \gamma_{\nu} (m_B) \gamma_{5}\left[\frac{\nsl_+}{\sqrt{2}}\phi^+_B(x_1) \right] }
    {(p_2-k_1+l)^2(k_1-k_2-l)^2(k_2+l)^2[(p_1-k_1+l)^2-m^2_b] l^2}  \bigg{\}} \non
    &=&   \frac{1}{8} \phi^{(1)}_{B,d} \otimes G^{(0)}_{aL}(\xi_1,x_2),   \label{eq:GaLf-i}\\
    %%%%%%
    G^{(1)}_{aL,2j} &=&  -\frac{eg^4Tr[T^aT^cT^aT^c]}{2N_c}  \non
    && \times \mathbf{Tr} \bigg{\{} \frac{\gamma^{\nu} (\ksl_1-\lsl) \gamma^{\alpha} (\ksl_2-\lsl) \gamma_{\nu} \left[M_{\rho} \esl_{2L}
    \phi_{\rho}(x_{2})\right] \gamma_{\alpha} (\psl_2-\ksl_1 ) \gamma_{\mu} (m_B) \gamma_{5}\left[\frac{\nsl_+}{\sqrt{2}}\phi^+_B(x_1) \right] }
    {(p_2-k_1+l)^2(k_1-k_2)^2(k_1-l)^2(k_2-l)^2 l^2}  \bigg{\}} \non
    &=&  - \frac{1}{8} \phi^{(1)}_{B,e} \otimes G^{(0)}_{aL}(x_1,x_2),  \label{eq:GaLf-j} \\
    %%%%%%
    G^{(1)}_{aL,2k} &=&  \frac{eg^4Tr[T^aT^cT^aT^c]}{2N_c}  \non
    && \times \mathbf{Tr} \bigg{\{} \frac{\gamma^{\nu} (\ksl_1-\lsl) \gamma^{\alpha} \left[M_{\rho} \esl_{2L} \phi_{\rho}(x_{2})\right]
    \gamma_{\nu}(\psl_2-\ksl_2-\lsl) \gamma_{\alpha} (\psl_2-\ksl_1 ) \gamma_{\mu} (m_B) \gamma_{5}\left[\frac{\nsl_+}{\sqrt{2}}\phi^+_B(x_1)\right]}
    {(p_2-k_1)^2(k_1-k_2-l)^2(p_2-k_2-l)^2(k_1-l)^2l^2}  \bigg{\}} \non
    &=&   \frac{1}{8} \phi^{(1)}_{B,e} \otimes G^{(0)}_{aL}(x_1,\xi_1,x_2),      \label{eq:GaLf-k}
	\eeq}
where the infrared divergence in Figs.~\ref{fig:2}(f,g) are suppressed by kinematics, and we briefly set them to zero for we concentrate on
infrared divergence only in factorization theorem here. All the diagrams Figs.~\ref{fig:2}(d-g) generate no soft divergence, for the reason
that soft divergence only occurs when the additional gluon bridge the initial and final quarks as Figs.~\ref{fig:2}(h-k).
Then these diagrams Figs.~\ref{fig:2}(h-k) will generate the soft divergence as well as the collinear divergence. But fortunately, we know that
these soft divergence will cancel between Fig.~\ref{fig:2}(h) and Fig.~\ref{fig:2}(i), Fig.~\ref{fig:2}(j) and Fig.~\ref{fig:2}(k) by a simple deformation
of propagators.
This conclusion is also supported by analytic calculations for quark diagrams as shown in Refs.~\cite{PQCD-NLO-Cheng,PQCD-NLO-Li,Zhang-Hua}.

We will sum up all the irreducible diagrams to correct the color factors and ensure the gauge invariance.
All the diagrams Figs.~\ref{fig:2}(d,f,h,i) with the additional gluon emitted from the up-quark lines of $B$ meson give the result:
{\small
	\beq
	G^{(1)}_{up,aL}(x_1,x_2) = \phi^{(1)}_{B,d}(x_1,\xi_1) \otimes \left[G^{(0)}_{aL}(\xi_1,x_1,x_2)-G^{(0)}_{aL}(\xi_1,x_2) \right].
	\label{eq:aLup}	
    \eeq}
The diagrams Figs.~\ref{fig:2}(e,g,j,k) with the additional gluon emitted from the down-quark lines of $B$ meson give the result:
{\small
	\beq
	G^{(1)}_{down,aL}(x_1,x_2) = \phi^{(1)}_{B,e}(x_1,\xi_1) \otimes \left[G^{(0)}_{aL}(x_1,x_2)-G^{(0)}_{aL}(x_1,\xi_1,x_2) \right].
	\label{eq:aLdown}	
	\eeq}
We take the similar representation as our previous work \cite{prove}, the function $G^{(0)}_{aL}(\xi_1,x_1,x_2)$, $G^{(0)}_{aL}(x_1,\xi_1,x_2)$ and
$ G^{(0)}_{aL}(x_1,\xi_1,x_2)$ in Eqs.~(\ref{eq:aLup},\ref{eq:aLdown}) represent the different forms of LO hard kernel with gluon momentum.
The different location of $\xi_1$ in vertex represents tiny difference of types as gluon momentum flowing into hard kernel.
{\small
	\beq
	G^{(0)}_{aL}(\xi_1,x_1,x_2) &=& -\frac{ieg^2C_F}{2} \mathbf{Tr}
	\bigg{\{} \frac{\gamma^{\alpha}[M_{\rho} \esl_{2L} \phi_{\rho}(x_{2})] \gamma_{\alpha} ( \psl_2 - \ksl_1 + \lsl) \gamma_{\mu} ( m_B)
\gamma_{5}\left[\frac{\nsl_+}{\sqrt{2}}\phi^+_B(x_1) \right]}
	{(p_2-k_1+l)^2(k_1-k_2)^2} \bigg{\}}, \label{eq:GaLxi-1}  \\
	G^{(0)}_{aL}(x_1,\xi_1,x_2) &=& -\frac{ieg^2C_F}{2} \mathbf{Tr}
	\bigg{\{} \frac{\gamma^{\alpha}[M_{\rho} \esl_{2L} \phi_{\rho}(x_{2})] \gamma_{\alpha} ( \psl_2 - \ksl_1) \gamma_{\mu} ( m_B)
\gamma_{5}\left[\frac{\nsl_+}{\sqrt{2}}\phi^+_B(x_1) \right]}
	{(p_2-k_1)^2(k_1-k_2-l)^2} \bigg{\}},  \label{eq:GaLxi-2}\\
	G^{(0)}_{aL}(\xi_1,x_2) &=& -\frac{ieg^2C_F}{2} \mathbf{Tr}
	\bigg{\{} \frac{\gamma^{\alpha}[M_{\rho} \esl_{2L} \phi_{\rho}(x_{2})] \gamma_{\alpha} ( \psl_2 - \ksl_1 + \lsl) \gamma_{\mu} ( m_B)
\gamma_{5}\left[\frac{\nsl_+}{\sqrt{2}}\phi^+_B(x_1) \right]}
	{(p_2-k_1+l)^2(k_1-k_2-l)^2} \bigg{\}}.  	\label{eq:GaLxi-3}
	\eeq}

\begin{figure}[htbp]
	\centering
	\includegraphics[width=4.5in]{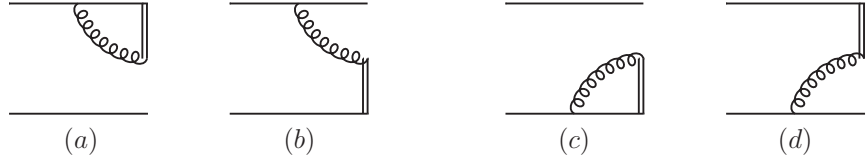}
	\caption{The effective diagrams for the NLO initial $B$ meson. }
	\label{fig:eff}
\end{figure}

After the cancellation of soft divergence, the remaining collinear divergence will be absorbed into those effective wave functions of the mesons involved.
The effective diagrams for the NLO initial $B$ meson are shown as Fig.~\ref{fig:eff}, where the Figs.~\ref{fig:eff}(a,c) represent diagrams with no
gluon momentum flows into LO hard
kernel, relatively, Figs.~\ref{fig:eff}(b,d) represent diagrams with the gluon momentum flowing into LO hard kernel. The nonlocal hadronic matrix element
for  the B meson wave function can be written as:
{\small
\beq
\Phi^{(1)}_{B}&=&\frac{1}{2N_c P^+_1} \int \frac{dy^-}{2\pi} e^{-i x p^+_1 y^-}
<0| \overline{q}(y^-) \Gamma  (-i g_s)  \int^{y^-}_{0} dzv
A(z\nu) q(0)|h_{\nu} \bar{q}(p_1))>, \label{eq:Bflow}
\eeq}
where $A(z \nu)$ and $h_{\nu}$ represent the gauge field and the effective heavy-quark field respectively, and $\Gamma$ represents gamma matrix decided by
specific part of twists, which can be chosen as $\gamma_{5} \gamma^- /2$ here in this channel.

\begin{figure}[htbp]
	\centering
	\includegraphics[width=4.5in]{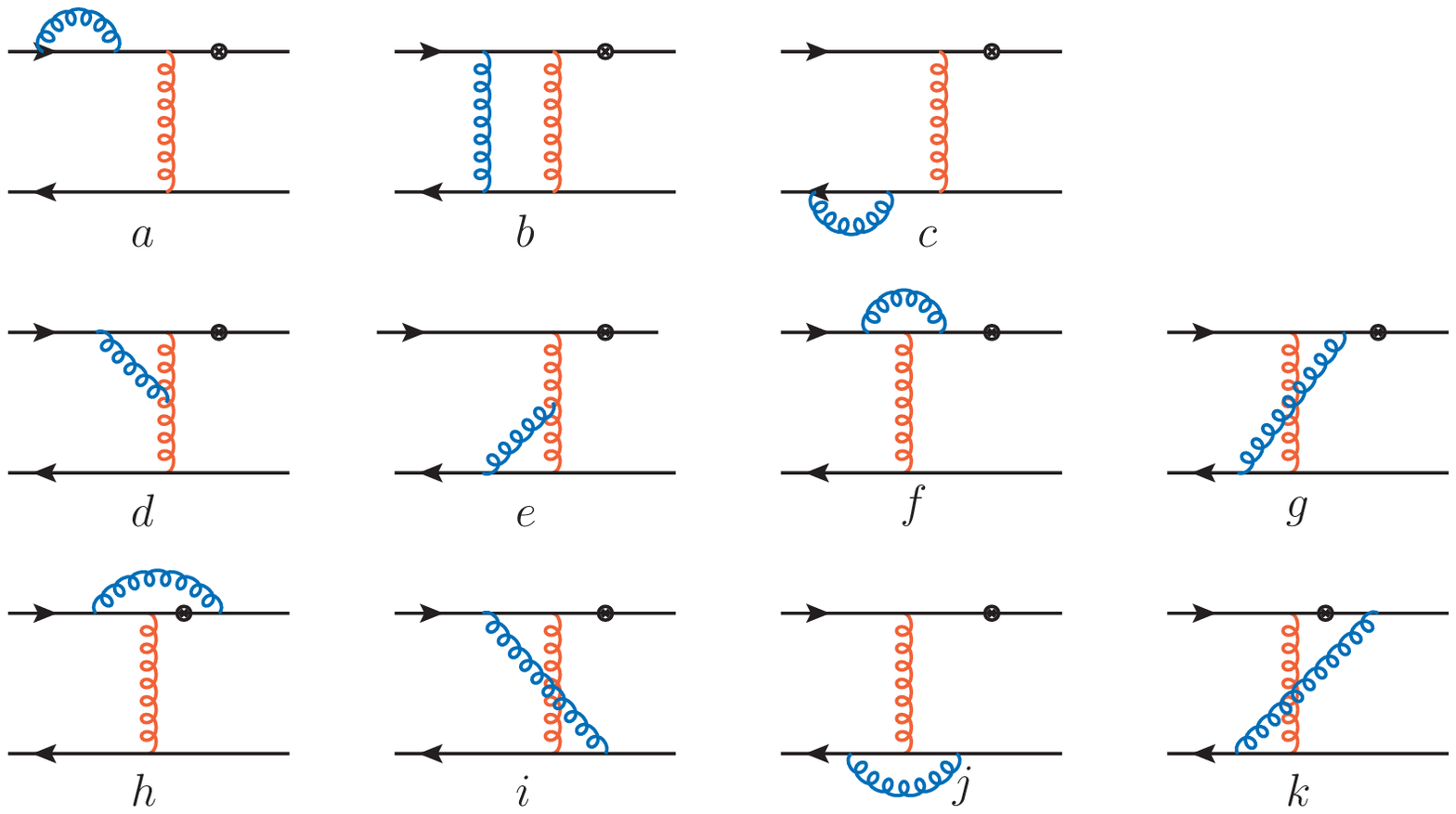}
	\caption{The quark diagrams for NLO corrections to Fig.~\ref{fig:1}(b) with additional gluon emitted from the quark(antiquark) lines of initial B meson.  }
	\label{fig:4}
\end{figure}

Now we consider the NLO corrections to one of the structures of Fig.~\ref{fig:1}(b). The first part of the LO amplitude $H^{(0)}_{bL}$ as given in Eq.~\ref{eq:Hbl} can be written in the form of
{\small
    \beq
    G^{(0)}_{bL,1} &=& \frac{-ieg^2C_Fm_B}{2}\mathbf{Tr} \left[ \frac{ \gamma^{\alpha}M_{\rho} \esl_{2L} \phi_{\rho}(x_{2}) \gamma_{\mu}
    \ksl_2 \gamma_{\alpha} \gamma_{5} \frac{\nsl+}{\sqrt{2}}\phi^+_B(x_1)}
    {[(p_1-k_2)^2-m^2_b](k_1-k_2)^2}\right].
    \label{eq:Gbl}
    \eeq}

We retain the $\gamma^{\alpha}$ matrix which should be $\gamma_{\perp}$ in Eq.~(\ref{eq:Gbl}) for convenience. The factorization about NLO
corrections to this part of the LO hard kernel gives the results as follows. The NLO $B$ meson wave function containing infrared divergence is of the form
as defined in Eq.~(\ref{eq:WFbL1de}). Each sub-diagram of this channel contributes without dynamical forbidden.
While, the change of weak vertex's location leads to different color structure for Figs.~\ref{fig:4}(f,g) comparing with those in Fig.~\ref{fig:2}.

{\small
	\beq
	G^{(1)}_{bL1,2d} &=& eg^4
	\mathbf{Tr} \bigg{\{} \frac{\gamma^{\alpha}\left[M_{\rho} \esl_{2L} \phi_{\rho}(x_{2})\right] \gamma_{\mu} \ksl_2
\gamma^{\beta} (\psl_1-\ksl_1+\lsl+m_b) \gamma^{\gamma} (m_B) \gamma_{5}\left[\frac{\nsl_+}{\sqrt{2}}\phi^+_B(x_1)\right] F_{\alpha \beta \gamma} }
	{(k_1-k_2)^2(k_1-k_2-l)^2[(p_1-k_2)^2-m^2_b][(p_1-k_1+l)^2-m^2_b]l^2}  \bigg{\}}   \non
	&=& -\frac{9}{8} \phi^{(1)}_{B,d} \otimes \left[G^{(0)}_{bL,1}(x_1,x_2)-G^{(0)}_{bL,1}(x_1,\xi_1,x_2) \right] , 	\label{eq:GbL1-2d}\\
	%%%%%
	G^{(1)}_{bL1,2e} &=& -eg^4
	\mathbf{Tr} \bigg{\{} \frac{\gamma^{\alpha}(\ksl_1-\lsl) \gamma^{\beta} \left[M_{\rho} \esl_{2L} \phi_{\rho}(x_{2})\right]
\gamma_{\mu} \ksl_2 \gamma^{\gamma}  (m_B)\gamma_{5}\left[\frac{\nsl_+}{\sqrt{2}}\phi^+_B(x_1)\right] F_{\alpha \beta \gamma} }
	{(k_1-k_2)^2(k_1-k_2-l)^2[(p_1-k_2)^2-m^2_b](k_1-l)^2l^2}  \bigg{\}}   \non
	&=& -\frac{9}{8} \phi^{(1)}_{B,e} \otimes \left[G^{(0)}_{bL,1}(x_1,x_2)-G^{(0)}_{bL,1}(x_1,\xi_1,x_2) \right],
	\label{eq:GbL1de}
\eeq}
with
{\small
\beq
	\phi^{(1)}_{B,d} &=& \frac{-ig^2C_F}{4} \frac{ \gamma^+\gamma_{5}  (\psl_1-\ksl_1+\lsl+m_b) \gamma^{\rho}
\gamma_{5}\gamma^- }{[(p_1-k_1+l)^2-m^2_b] l^2} \frac{\nu_{\rho}}{\nu \cdot l} , 	\label{eq:phi1b-d1}\\
	%%%%
	\phi^{(1)}_{B,e} &=& \frac{ig^2C_F}{4} \frac{ \gamma_{5}\gamma^- \gamma^{\rho} (\ksl_1-\lsl)  \gamma^+
\gamma_{5} }{(k_1-l)^2 l^2} \frac{\nu_{\rho}}{\nu \cdot l}.
	\label{eq:WFbL1de}
	\eeq}
and
{\small
	\beq
	G^{(1)}_{bL1,2f} &=& \frac{eg^4}{9}
	\mathbf{Tr} \bigg{\{} \frac{\gamma^{\alpha} \left[M_{\rho} \esl_{2L} \phi_{\rho}(x_{2})\right] \gamma_{\mu} \ksl_2
\gamma^{\nu} (\psl_1-\ksl_2 + \lsl + m_b) \gamma_{\alpha} (\psl_1-\ksl_1+\lsl+m_b) \gamma_{\nu}  (m_B) \gamma_{5}
\left[\frac{\nsl_+}{\sqrt{2}}\phi^+_B(x_1) \right] }
	{[(p_1-k_2)^2-m^2_b](k_1-k_2)^2[(p_1-k_2+l)^2-m^2_b][(p_1-k_1+l)^2-m^2_b] l^2}  \bigg{\}} \non
	&=& \frac{1}{8} \phi^{(1)}_{B,d} \otimes \left[G^{(0)}_{bL,1}(x_1,x_2)-G^{(0)}_{bL,1}(\xi_1,x_1,x_2) \right] , 	\label{eq:GbL1-2f}  \\
	%%%%%%
	G^{(1)}_{bL1,2g} &=& -\frac{eg^4}{9}
	\mathbf{Tr} \bigg{\{} \frac{\gamma^{\nu} (\ksl_1 - \lsl) \gamma^{\alpha} \left[M_{\rho} \esl_{2L} \phi_{\rho}(x_{2})\right]
\gamma_{\mu} \ksl_2 \gamma_{\nu} (\psl_1-\ksl_2-\lsl+m_b)  \gamma_{\alpha}  (m_B)\gamma_{5}\left[\frac{\nsl_+}{\sqrt{2}}\phi^+_B(x_1) \right] }
	{[(p_1-k_2)^2-m^2_b](k_1-k_2-l)^2[(p_1-k_2-l)^2-m^2_b](k_1-l)^2 l^2}  \bigg{\}} \non
	&=& -\frac{1}{8} \phi^{(1)}_{B,e}  \otimes \left[G^{(0)}_{bL,1}(x_1,\xi_1,x_2)-G^{(0)}_{bL,1}(\xi_1,x_2) \right]  ,  	\label{eq:GbL1-2g}
\eeq}
	%%%%%%
{\small
\beq
	G^{(1)}_{bL1,2h} &=&  \frac{eg^4}{9}
	\mathbf{Tr} \bigg{\{} \frac{\gamma^{\alpha} \left[M_{\rho} \esl_{2L} \phi_{\rho}(x_{2})\right] \gamma^{\nu}  (\psl_2-\ksl_2+\lsl)
\gamma_{\mu} (\ksl_2 -\lsl ) \gamma_{\alpha}  (\psl_1-\ksl_1+\lsl+m_b) \gamma_{\nu} (m_B) \gamma_{5}\left[\frac{\nsl_+}{\sqrt{2}}\phi^+_B(x_1) \right] }
	{[(p_1-k_2+l)^2-m^2_b](k_1-k_2)^2(p_2-k_2+l)^2[(p_1-k_1+l)^2-m^2_b] l^2}  \bigg{\}} \non
	&=&   \frac{1}{8} \phi^{(1)}_{B,d} \otimes G^{(0)}_{bL,1}(\xi_1,x_1,x_2), 	\label{eq:GbL1-2h} \\
	%%%%%%
	G^{(1)}_{bL1,2i} &=&  -\frac{eg^4}{9}
	\mathbf{Tr} \bigg{\{} \frac{\gamma^{\alpha} (\ksl_2+\lsl) \gamma^{\nu} \left[M_{\rho} \esl_{2L} \phi_{\rho}(x_{2})\right]
\gamma_{\mu} \ksl_2 \gamma_{\alpha} (\psl_1-\ksl_1+\lsl+m_b) \gamma_{\nu} (m_B) \gamma_{5}\left[\frac{\nsl_+}{\sqrt{2}}\phi^+_B(x_1) \right] }
	{[(p_1-k_2)^2-m^2_b](k_1-k_2+l)^2(k_2+l)^2[(p_1-k_1+l)^2-m^2_b] l^2}  \bigg{\}} \non
	&=&   -\frac{1}{8} \phi^{(1)}_{B,d} \otimes G^{(0)}_{bL,1}(x_1,\xi_1,x_2) , 	\label{eq:GbL1-2i}
\eeq }
{\small
\beq
G^{(1)}_{bL1,2j} &=&  -\frac{eg^4}{9}
	\mathbf{Tr} \bigg{\{} \frac{\gamma^{\nu} (\ksl_1-\lsl) \gamma^{\alpha} (\ksl_2-\lsl) \gamma_{\nu} \left[M_{\rho} \esl_{2L}
\phi_{\rho}(x_{2})\right] \gamma_{\mu} \ksl_2 \gamma_{\alpha} (m_B) \gamma_{5}\left[\frac{\nsl_+}{\sqrt{2}}\phi^+_B(x_1) \right] }
	{[(p_1-k_2)^2-m^2_b](k_1-k_2)^2(k_1-l)^2(k_2-l)^2 l^2}  \bigg{\}} \non
	&=&   \frac{1}{8} \phi^{(1)}_{B,e} \otimes G^{(0)}_{bL,1}(x_1,x_2), 	\label{eq:GbL1-2j} \\
	%%%%%%
	G^{(1)}_{bL1,2k} &=&  \frac{eg^4}{9}
	\mathbf{Tr} \bigg{\{} \frac{\gamma^{\nu} (\ksl_1-\lsl) \gamma^{\alpha} \left[M_{\rho} \esl_{2L} \phi_{\rho}(x_{2})\right]
\gamma_{\nu}(\psl_2-\ksl_2-\lsl) \gamma_{\mu} (\ksl_2+\lsl) \gamma_{\alpha} (m_B) \gamma_{5}\left[\frac{\nsl_+}{\sqrt{2}}\phi^+_B(x_1)\right]}
	{[(p_1-k_2-l)^2-m^2_b](k_1-k_2-l)^2(p_2-k_2-l)^2(k_1-l)^2l^2}  \bigg{\}} \non
	&=&   -\frac{1}{8} \phi^{(1)}_{B,e} \otimes G^{(0)}_{bL,1}(\xi_1,x_2).	\label{eq:GbL1-2k}
	\eeq}

We also sum up all the contributions from the diagrams with additional gluon emitted from up-quark (down-quark) lines of initial $B$ meson
of this channel and the results are of the following form:
{\small
	\beq
	G^{(1)}_{up,bL1}(x_1,x_2) = -\phi^{(1)}_{B,d}(x_1,\xi_1) \otimes \left[G^{(0)}_{aL}(x_1,x_2)-G^{(0)}_{aL}(x_1,\xi_1,x_2) \right],
	\label{eq:bL1up} \non
	G^{(1)}_{down,bL1}(x_1,x_2) = -\phi^{(1)}_{B,e}(x_1,\xi_1) \otimes \left[G^{(0)}_{aL}(x_1,x_2)-G^{(0)}_{aL}(x_1,\xi_1,x_2) \right].
	\label{eq:bL1down}	
	\eeq}
The minus sign here is caused by the $-\ksl_2$ in gluon propagator at LO.  Although the Figs.~\ref{fig:4}(f,g) now give the different color structures,
the summation of all the sub-diagrams still give the right color factor which we can also be obtained from the effective diagrams directly. As we did
in $G^{1}_{aL}$ or other terms, after the cancellation of soft divergence, the remaining collinear divergence can be factorized into NLO $B$ meson wave
function without any other theoretical  problems.

\subsection{$k_T$ factorization of NLO corrections to $B \to \rho$ transition}

\begin{figure}[htbp]
	\centering
	\includegraphics[width=5.5in]{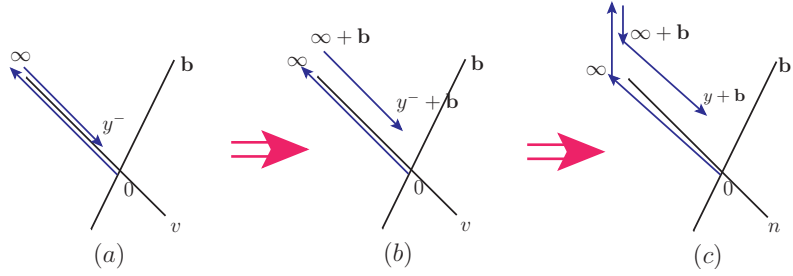}
	\vspace{-14.5cm}
	\caption{The graph for the deviation of the integral (Wilson line) by b in light cone coordinate for the two-parton meson wave function. }
	\label{fig:5}
\end{figure}

We pick up the transversal momentum $``k_T"$ in $k_T$ factorization frame to remove the endpoint singularity emerged in small $x$ region.
The Sudakov factor $e^{-S(t)}$ is introduced  \cite{ppnp51-85} in order  to suppress the collinear divergence by resummation of the large logarithmic
terms which $k_T$ is concerned.

As we proposed in $k_T$ factorization frame, the $k_T$ works only in small $x$ region, which indicates that we have the hierarchy $k^2_{iT} \ll k_1 \cdot k_2$.
Comparing with the expressions as shown in Eqs.~(\ref{eq:Hal}) and (\ref{eq:Hbt}), we can also keep the same forms of LO hard kernel in $k_T$ factorization frame.
After the inclusion of the NLO corrections, the only difference we should pay attention to is the transverse momentum $\mathbf{l_T}$ in propagators which can be understood
as a small momentum shift.

The ratio term $\nu_{\rho} / {\nu \cdot l}$ in Eq.~(\ref{eq:WFaLde}) represents the Feynman rule related to Wilson line.
Considering the $\mathbf{l_T}$, we rewrite
the hadronic matrix element in Eq.~(\ref{eq:Bflow}) containing Wilson line through Fourier transformation for the gauge field form from
$A(z \nu)$ to $\widetilde{A}(l)$:
\beq
\int^{\infty}_{0} dz\; v\cdot A(zv) & \to & \int^{\infty}_{0} dz\int dl ~e^{iz(v \cdot l + i\epsilon)}~v\cdot \widetilde{A}(l) \non
&\to& i \int dl ~\frac{v_\rho}{v\cdot l} \widetilde{A}^\rho(l), \label{eq:feyBfow1}\\
\int^{y^-}_{\infty} dz\; v\cdot A(zv)
& \to & \int^{y^-}_{0} dz \int dl ~e^{[iz(v \cdot l + i\epsilon)-i\mathbf{l}_T \cdot \mathbf{b}]}~v\cdot \widetilde{A}(l)\non
& \to & -i \int dl ~\frac{v_\rho}{v\cdot l} ~e^{[il^+y^- - i\mathbf{l}_T \cdot \mathbf{b}]}~\widetilde{A}^\rho(l),
\label{eq:feyBfow2}
\eeq	
The factor $e^{il^+y^-}$ will generate a delta function $\delta(\xi_1-x_1+\frac{l^+}{p_1})$ which describes
a momentum shift when the $l$ flowing into hard kernel.
 The other term $e^{i\mathbf{l}_T}$ represents a small transversal part flowing into the hard kernel. The NLO $B$ and $\rho$ meson wave
functions can be finally written as a general form:
\beq
\Phi^{(1)}_{B}(x_1,\xi_1;\mathbf{b_1})&=&\frac{1}{2N_c P^+_1} \int \frac{dy^-}{2\pi} \frac{d\mathbf{b_1}}{(2\pi)^2}
e^{-i x p^+_1 y^- + i \mathbf{k_{1T}} \cdot \mathbf{b_1}} \non
&&\cdot <0| \overline{q}(y^-)  \Gamma \cdot (-i g_s) \int^{y}_{0} dz\; n \cdot A(zn) q(0)|h_{\nu} B(p_1)>,
\label{eq:genB} \\
\Phi^{(1)}_{\rho}(x_2,\xi_2;\mathbf{b_2})=&&\frac{1}{2N_c P^-_2} \int \frac{dy^-}{2\pi} \frac{d\mathbf{b_2}}{(2\pi)^2}
e^{-i x p^+_2 y^- + i \mathbf{k_{2T}} \cdot \mathbf{b_2}} \non
&&\cdot <0|\overline{q}(y^-) \Gamma' \cdot (-i g_s)  \int^{y}_{0} dz \;n \cdot A(zn) q(0)|\rho(p_1)>.
\label{eq:genrho}
\eeq
where $\Gamma$ is the gamma matrix depending on the twist of the $B$ and $\rho$ wave functions.

The whole process of the integration and modification in Eqs.~(\ref{eq:feyBfow1},\ref{eq:feyBfow2}) can be understood by graphs
in Fig.~\ref{fig:5}. Both the Wilson lines in Fig.~\ref{fig:5}(a,b) parallel to the light cone coordinate will generate light cone singularity.
To avoid this singularity, we can slightly rotate the direction of Wilson line as in Fig.~\ref{fig:5}(c), then such a singularity can be regularized
by $n^2(n^2 \ne 0)$. The choice of number $n^2$ is scheme dependent, which has been demonstrated to be small \cite{li1302} and we can
simply set $n^2=1$ for convenience in calculation. While this rotation and the choose of $n^2$ will generate a pinched singularity during
the integration for Wilson lines. To solve this problem, the transverse-momentum-dependent(TMD) wave function with soft substraction factor
with a square root is introduced to the un-subtracted wave function \cite{ijmpcs4-85}, and a more elegant TMD wave function involves two
pieces of non-light-like Wilson links is proposed in Ref.\cite{jhep1506-013}. We pay more attention to the factorization of the collinear divergence
in this paper, more discussion on these TMD wave function will be given in future work.

 \section{SUMMARY}

In this paper, we investigate the $B \to \rho$ transition process and give the proof of factorization at NLO level in collinear factorization approach.
Because of the different structures in the initial $B$ meson and the final $\rho$ meson state,  three amplitudes with six parts of twists' forms for LO diagrams
as shown in Fig.~\ref{fig:1}(a,b) should be considered.  Also, both the corrections to initial state and finial state should be considered. The calculation
of all these six channels will be much complex, which suggests us an estimate for proportion for every channel in LO will be necessary
before numerical calculations done at the NLO level.

We have verified that with the right power counting for triple gluon vertex, the amplitudes of the NLO corrections to the LO hard kernel with an additional
gluon emitted from the initial $B$ meson or the final $\rho$ meson can be separated into a convolution of the NLO wave function with LO hard kernel.
The LO hard kernel here can be distinguished as two forms, with gluon momentum flowing into the hard kernel, or with no gluon momentum flowing into the
hard kernel. We then extend these results to $k_T$ factorization frame and the conclusion can still be stable by a momentum shift. The form of
nonlocal two quarks hadron matrices of NLO $B(\rho)$ meson are given in section.~\Rmnum{3}.
This work will also play an essential role for future numerical calculations of $B \to \rho$ transition form factors.

 \section*{ACKNOWLEDEMENT}

This work is supported by the National Natural Science
Foundation of China under Grants  No.~11775117 and 11235005,
and also by the Practice Innovation Program of Jiangsu Province under Grant No. KYCX18-1184.

%%%%%%%%%%%%%%%%%%%%%%%%%%%%%%%%%%%%%%%%%%%%%%%%%%%%%%%%%%%%%%%%%%%%%%%%%%%%%%%%                                       Appendix
%%%%%%%%%%%%%%%%%%%%%%%%%%%%%%%%%%%%%%%%%%%%%%%%%%%%%%%%%%%%%%%%%%%%%%%%%%%%%%%%

\begin{appendix}
	
\section{Factorization of the NLO  amplitudes }\label{sec:da}

In this Appendix, we present the explicit expressions for NLO corrections to the remaining LO amplitude $H^{0}_{bL}$  and  $H^{0}_{bT}$ .
The function $G^{(0)}_{bL2}$ corresponds to the second part of $H^{0}_{bL}$ as given in Eq.~(\ref{eq:Hbl}).
The function $G^{(0)}_{bT1}$, $G^{(0)}_{bT2}$ and $G^{(0)}_{bT3}$ come from the three components of
$H^{0}_{bT}$ as defined in Eq.~(\ref{eq:Hbt}): two for $\phi_B^+$ and one for $\phi_B^-$ component.
For the sake of clearness, we write these four functions in their  original non-reduced form:
{\small
\beq
G^{(0)}_{bL2} &=&-\frac{ieg^2C_F}{2} \mathbf{Tr} \left \{ \frac{\gamma^{\alpha} \left[M_{\rho} \esl_{2L} \phi_{\rho}(x_{2})\right] \gamma_{\mu} (m_b) \gamma_{\alpha} (\psl_1) \gamma_{5}\left[\frac{\nsl_-}{\sqrt{2}}\phi^-_B(x_1)\right]}
{[(p_1-k_2)^2-m^2_b](k_1-k_2)^2}\right \},     \label{eq:Gbl2} \\
	G^{(0)}_{bT1} &=&-\frac{ieg^2C_F}{2} \mathbf{Tr} \left \{[\frac{\gamma^{\alpha} \left[\esl_{2T} \psl_2 \phi^T_{\rho}(x_{2})\right] \gamma_{\mu} (\psl_1)
	\gamma_{\alpha} (\psl_1) \gamma_{5}\left[\frac{\nsl_+}{\sqrt{2}}\phi^+_B(x_1) \right] }
	{\Big[(p_1-k_2)^2-m^2_b\Big](k_1-k_2)^2} \right\},	
	\label{eq:GbT1} \\
	G^{(0)}_{bT2} &=&-\frac{ieg^2C_F}{2} \mathbf{Tr} \left \{\frac{\gamma^{\alpha} \left[\esl_{2T} \psl_2 \phi^T_{\rho}(x_{2})\right] \gamma_{\mu} (\psl_1)
	\gamma_{\alpha} (\psl_1) \gamma_{5}\left[\frac{\nsl_-}{\sqrt{2}}\phi^-_B(x_1) \right]}
	{\Big[(p_1-k_2)^2-m^2_b\Big](k_1-k_2)^2} \right\},	
	\label{eq:GbT2} \\
	G^{(0)}_{bT3} &=&-\frac{ieg^2C_F}{2} \mathbf{Tr} \left \{\frac{
	\gamma^{\alpha} \left[\esl_{2T} \psl_2 \phi^T_{\rho}(x_{2})\right] \gamma_{\mu} (m_b)
	\gamma_{\alpha} (m_B) \gamma_{5}\left[\frac{\nsl_+}{\sqrt{2}}\phi^+_B(x_1) \right]}
	{\Big[(p_1-k_2)^2-m^2_b\Big](k_1-k_2)^2} \right\}.	
	\label{eq:GbT3}
	\eeq}

\begin{figure}[htb]
	\centering
	\includegraphics[width=4.5in]{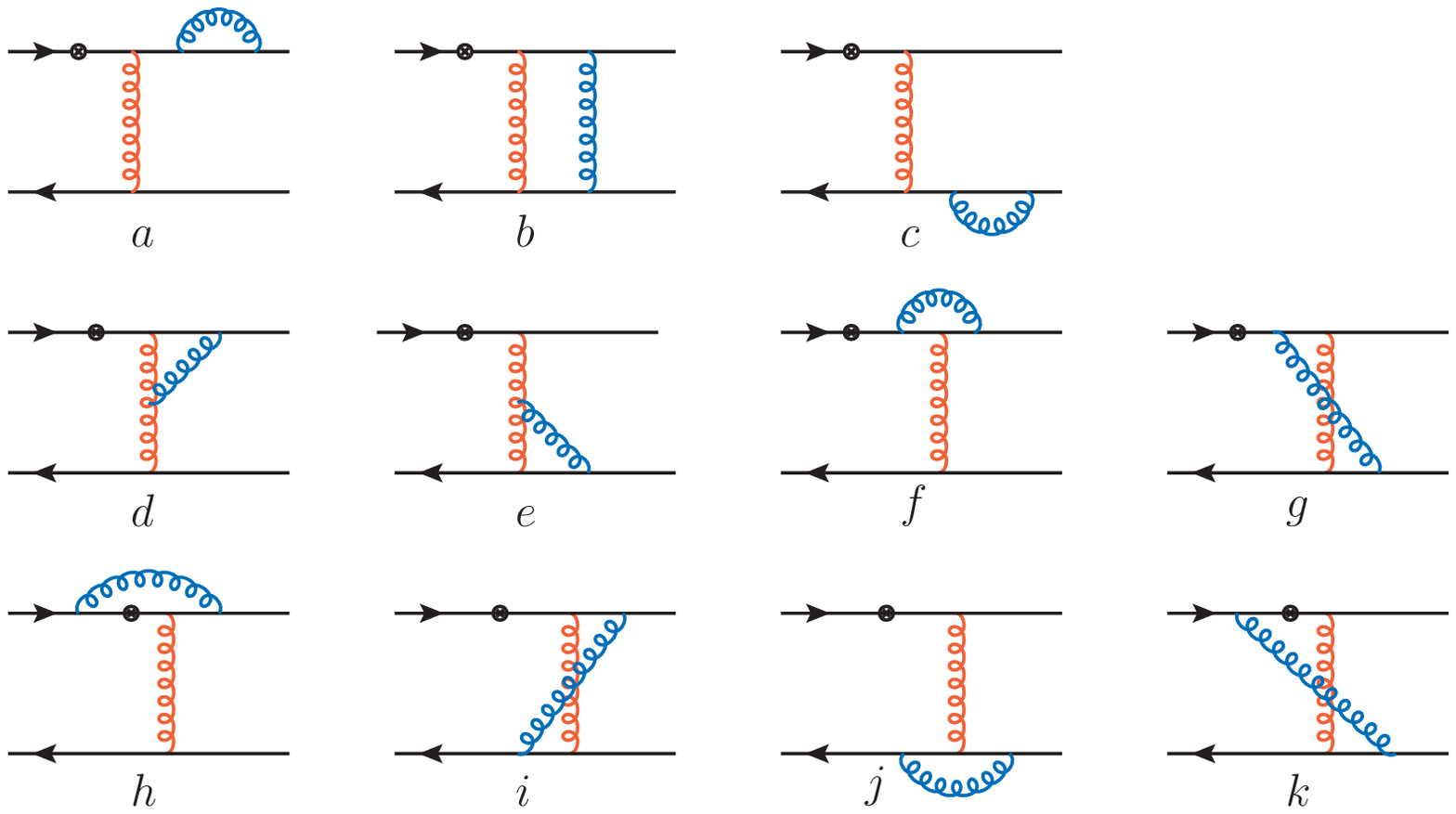}
	\caption{ The Feynman diagrams for NLO corrections to Fig.~\ref{fig:1}(a) with the additional gluon emitted from the quark (antiquark) lines of
the finial $\rho$ meson. }
	\label{fig:6}
\end{figure}

\begin{figure}[htb]
	\centering
	\includegraphics[width=4.5in]{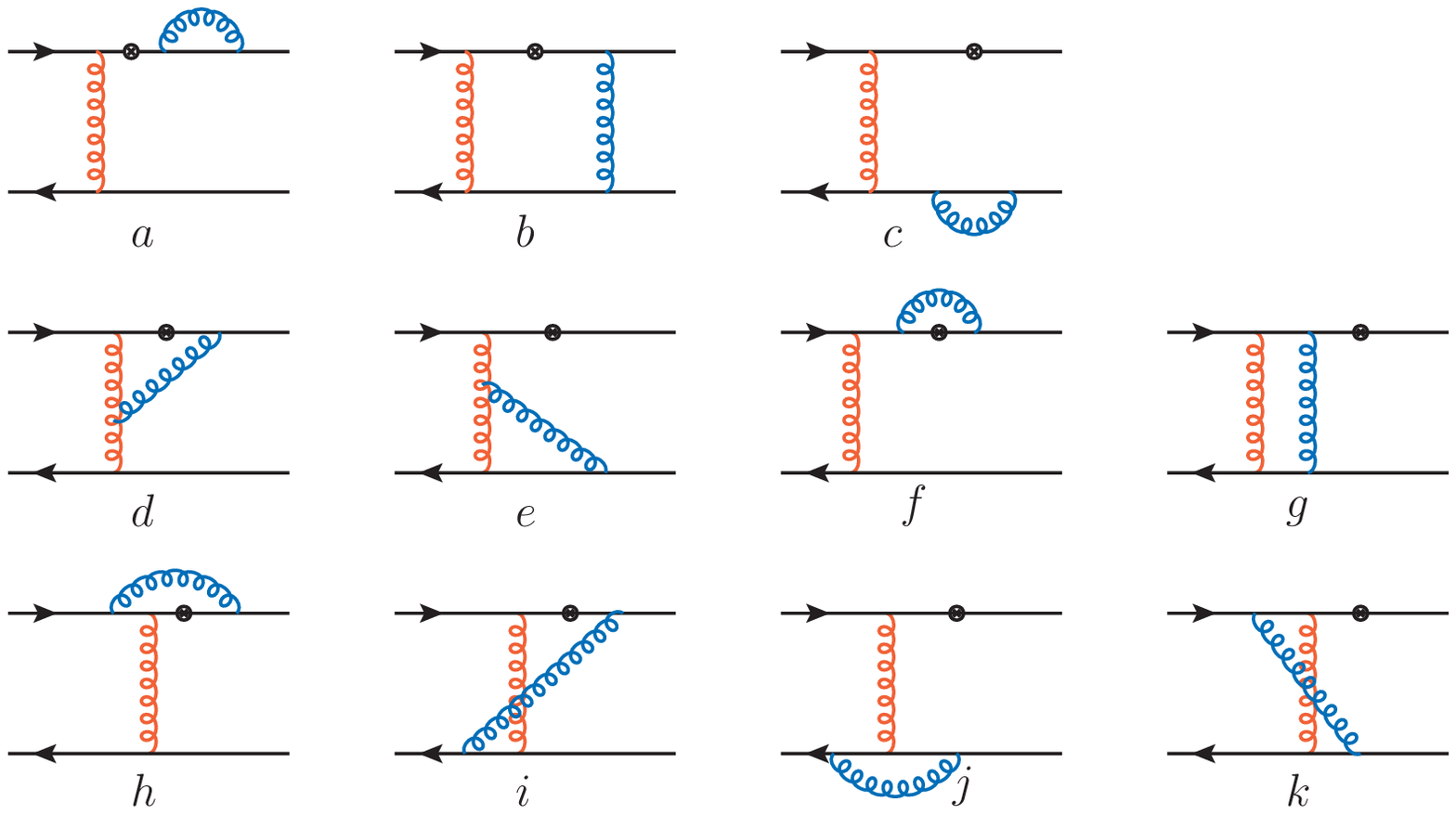}
	\caption{The Feynman diagrams for NLO corrections to Fig.~\ref{fig:1}(b) with the additional gluon emitted from the quark (antiquark) lines of
the finial $\rho$ meson.  }
	\label{fig:7}
\end{figure}

The NLO corrections with additional gluon emitted from the final $\rho$ meson are also supplied.
In Fig.~\ref{fig:6} and \ref{fig:7} , we show the Feynman diagrams for NLO corrections to Fig.~\ref{fig:1}(a) and Fig.~\ref{fig:1}(b)  respectively, where
the second blue gluon are emitted from the quark and antiquark lines of the finial $\rho$ meson.
Notice that the $G^{(0)}_{aL}$ and $G^{(0)}_{bL1}$ with the gluon emitted from the final $\rho$ meson
have the similar structure with the ones for $\rho \to \rho$ decays as defined in Refs.~\cite{prove},
so we here do not present them explicitly.

%%%%%%%%%%%%%%%%%%%%%%%%%%%%%%%%%%%%%%%%%%%%%%%%%%%%%%%%%%%%%%%%%%%%%%%%%%%%%%%

\subsection{The NLO amplitudes for $G^{(0)}_{bL2}$ }

The amplitudes for the NLO contributions from the irreducible diagrams to $G^{(0)}_{bL2}$ are the following:
{\small
	\beq
	G^{(1)}_{bL2,2d} &=& -eg^4
	\mathbf{Tr} \bigg \{ \frac{\gamma^{\alpha}\left[M_{\rho} \esl_{2L} \phi_{\rho}(x_{2})\right]
\gamma_{\mu} m_b  \gamma^{\beta} (\psl_1-\ksl_1+\lsl+m_b) \gamma^{\gamma} (\psl_1) \gamma_{5}
\left[\frac{\nsl_-}{\sqrt{2}}\phi^-_B(x_1)\right] F_{\alpha \beta \gamma} }
	{(k_1-k_2)^2(k_1-k_2-l)^2[(p_1-k_2)^2-m^2_b][(p_1-k_1+l)^2-m^2_b]l^2}  \bigg \}    \non
	&=& \frac{9}{8} \phi^{(1)}_{B,d} \otimes \left[G^{(0)}_{bL,2}(x_1,x_2)-G^{(0)}_{bL,2}(x_1,\xi_1,x_2) \right] , \\
	%%%%%
	G^{(1)}_{bL2,2e} &=& -eg^4
\times \mathbf{Tr} \bigg \{ \frac{\gamma^{\alpha}(\ksl_1-\lsl) \gamma^{\beta} \left[M_{\rho} \esl_{2L} \phi_{\rho}(x_{2})\right]
\gamma_{\mu} m_b \gamma^{\gamma}  (\psl_1)\gamma_{5}\left[\frac{\nsl_-}{\sqrt{2}}\phi^-_B(x_1)\right] F_{\alpha \beta \gamma} }
    {(k_1-k_2)^2(k_1-k_2-l)^2[(p_1-k_2)^2-m^2_b](k_1-l)^2l^2}  \bigg \}   \non
    && \sim 0. 	\label{eq:GbL2de}
	\eeq}
The contribution of $G^{(1)}_{bL2,2e}$ is suppressed by dynamics and we simply set it $0$ here. For other contributions we have:

{\small
\beq
G^{(1)}_{bL2,2f} &=& \frac{eg^4}{9}
\mathbf{Tr} \bigg \{ \frac{\gamma^{\alpha} \left[M_{\rho} \esl_{2L} \phi_{\rho}(x_{2})\right] \gamma_{\mu} m_b  \gamma^{\nu}
(\psl_1-\ksl_2 + \lsl + m_b) \gamma_{\alpha} (\psl_1-\ksl_1+\lsl+m_b) \gamma_{\nu}  (\psl_1) \gamma_{5}\left[\frac{\nsl_-}{\sqrt{2}}\phi^-_B(x_1)\right] }
{[(p_1-k_2)^2-m^2_b](k_1-k_2)^2[(p_1-k_2+l)^2-m^2_b][(p_1-k_1+l)^2-m^2_b] l^2}  \bigg \}, \non
&& \sim 0,  \\
	%%%%%%
G^{(1)}_{bL2,2g} &=& -\frac{eg^4}{9}
\mathbf{Tr} \bigg \{ \frac{\gamma^{\nu} (\ksl_1 - \lsl) \gamma^{\alpha} \left[M_{\rho} \esl_{2L} \phi_{\rho}(x_{2})\right]  \gamma_{\mu} m_b
\gamma_{\nu} (\psl_1-\ksl_2+\lsl+m_b)  \gamma_{\alpha}  (\psl_1)\gamma_{5}\left[\frac{\nsl_-}{\sqrt{2}}\phi^-_B(x_1)\right] }
{[(p_1-k_2)^2-m^2_b](k_1-k_2-l)^2[(p_1-k_2+l)^2-m^2_b](k_1-l)^2 l^2}  \bigg \} \non
&& \sim 0, \\
	%%%%%%
G^{(1)}_{bL2,2h} &=&  \frac{eg^4}{9}
 \mathbf{Tr} \bigg \{ \frac{\gamma^{\alpha} \left[M_{\rho} \esl_{2L} \phi_{\rho}(x_{2})\right] \gamma^{\nu}
 (\psl_2-\ksl_2+\lsl) \gamma_{\mu} (m_b +\lsl ) \gamma_{\alpha}  (\psl_1-\ksl_1+\lsl+m_b) \gamma_{\nu} (\psl_1) \gamma_{5}
  \left[\frac{\nsl_-}{\sqrt{2}}\phi^-_B(x_1)\right] }
{[(p_1-k_2+l)^2-m^2_b](k_1-k_2)^2(p_2-k_2+l)^2[(p_1-k_1+l)^2-m^2_b] l^2}  \bigg \} \non
&=&   -\frac{1}{8} \phi^{(1)}_{B,d} \otimes G^{(0)}_{bL,1}(x_1,x_2) , \\
	%%%%%%
G^{(1)}_{bL2,2i} &=&  -\frac{eg^4}{9}
\mathbf{Tr} \bigg \{ \frac{\gamma^{\alpha} (\ksl_2 +\lsl) \gamma^{\nu} \left[M_{\rho} \esl_{2L} \phi_{\rho}(x_{2})\right]
\gamma_{\mu} m_b \gamma_{\alpha} (\psl_1-\ksl_1+\lsl+m_b) \gamma_{\nu} (\psl_1) \gamma_{5}\left[\frac{\nsl_-}{\sqrt{2}}\phi^-_B(x_1)\right] }
{[(p_1-k_2)^2-m^2_b](k_1-k_2+l)^2(k_2+l)^2[(p_1-k_1+l)^2-m^2_b] l^2}  \bigg \} \non
&=&   \frac{1}{8} \phi^{(1)}_{B,d} \otimes G^{(0)}_{bL,1}(x_1,\xi_1,x_2) , \\
	%%%%%%
G^{(1)}_{bL2,2j} &=&  \frac{eg^4}{9}
\mathbf{Tr} \bigg \{ \frac{\gamma^{\nu} (\ksl_1-\lsl) \gamma^{\alpha} (\ksl_2-\lsl) \gamma_{\nu} \left[M_{\rho} \esl_{2L} \phi_{\rho}(x_{2})\right]
\gamma_{\mu} m_b  \gamma_{\alpha} (\psl_1) \gamma_{5}\left[\frac{\nsl_-}{\sqrt{2}}\phi^-_B(x_1)\right] }
{[(p_1-k_2)^2-m^2_b](k_1-k_2)^2(k_1-l)^2(k_2-l)^2 l^2}  \bigg  \} \non
 && \sim 0,  \\
	%%%%%%
G^{(1)}_{bL2,2k} &=&  -\frac{eg^4}{9}
\mathbf{Tr} \bigg \{ \frac{\gamma^{\nu} (\ksl_1-\lsl) \gamma^{\alpha} \left[M_{\rho} \esl_{2L} \phi_{\rho}(x_{2})\right]
\gamma_{\nu}(\psl_2-\ksl_2-\lsl) \gamma_{\mu} (m_b -\lsl) \gamma_{\alpha} (\psl_1) \gamma_{5}\left[\frac{\nsl_-}{\sqrt{2}}\phi^-_B(x_1)\right]}
{[(p_1-k_2-l)^2-m^2_b](k_1-k_2-l)^2(p_2-k_2-l)^2(k_1-l)^2l^2}  \bigg \} \non
&& \sim 0, \label{eq:GbL2f-k}
\eeq}
with
{\small
\beq
\phi^{(1)}_{B,d} &=& \frac{-ig^2C_F}{8} \frac{ \gamma^-\gamma_{5}\gamma^+  (\psl_1-\ksl_1+\lsl+m_b) \gamma^{\rho} \gamma^-
\gamma_{5}\gamma^+ }{[(p_1-k_1+l)^2-m^2_b] l^2} \frac{\nu_{\rho}}{\nu \cdot l}. \label{eq:WFbL2d}
\eeq}

The summation over above contributions gives the result:
{\small
	\beq
	G^{(1)}_{up,bL2}(x_1,x_2) = \phi^{(1)}_{B,d}(x_1,\xi_1) \otimes \left[G^{(0)}_{aL}(x_1,x_2)-G^{(0)}_{aL}(x_1,\xi_1,x_2) \right].
	\label{eq:bL2up}
\eeq}

%%%%%%%%%%%%%%%%%%%%%%%%%%%%%%%%%%%%%%%%%%%%%%%%%%%%%%%%%%%%%%%%%%%%%%%%%%%%%%%
%%%%%%%%%%%%%%%%%%%%%%%%%%%%%%%%%%%%%%%%%%%%%%%%%%%%%%%%%%%%%%%%%%%%%%%%%%%%%%%

\subsection{The NLO amplitudes for $G^{(0)}_{bT1}$ }

The amplitudes for the NLO contributions from the irreducible diagrams to $G^{(0)}_{bT1}$ are the following:
{\small
	\beq
G^{(1)}_{bT1,2d} &=& -eg^4
\mathbf{Tr} \bigg \{ \frac{\gamma^{\alpha}\left[\esl_{2T} \psl_2 \phi^T_{\rho}(x_{2})\right] \gamma_{\mu} \psl_1
\gamma^{\beta} (\psl_1-\ksl_1+\lsl+m_b) \gamma^{\gamma} (\psl_1) \gamma_{5}\left[\frac{\nsl_+}{\sqrt{2}}\phi^+_B(x_1)\right] F_{\alpha \beta \gamma} }
{(k_1-k_2)^2(k_1-k_2-l)^2[(p_1-k_2)^2-m^2_b][(p_1-k_1+l)^2-m^2_b]l^2}  \bigg \}   \non
&=& \frac{9}{8} \phi^{(1)}_{B,d} \otimes \left[G^{(0)}_{bT,1}(x_1,x_2)-G^{(0)}_{bT,1}(x_1,\xi_1,x_2) \right],  \\
	%%%%%
G^{(1)}_{bT1,2e} &=& eg^4
\mathbf{Tr} \bigg \{  \frac{\gamma^{\alpha}(\ksl_1-\lsl) \gamma^{\beta} \left[\esl_{2T} \psl_2 \phi^T_{\rho}(x_{2})\right]
\gamma_{\mu} \psl_1 \gamma^{\gamma}  (\psl_1)\gamma_{5}\left[\frac{\nsl_+}{\sqrt{2}}\phi^+_B(x_1)\right] F_{\alpha \beta \gamma} }
{(k_1-k_2)^2(k_1-k_2-l)^2[(p_1-k_2)^2-m^2_b](k_1-l)^2l^2}  \bigg \}   \non
&=& \frac{9}{8} \phi^{(1)}_{B,e} \otimes \left[G^{(0)}_{bT,1}(x_1,x_2)-G^{(0)}_{bT,1}(x_1,\xi_1,x_2) \right],
\label{eq:GbT1de}
\eeq}
{\small
\beq
G^{(1)}_{bT1,2f} &=& \frac{eg^4}{9}
\mathbf{Tr} \bigg \{  \frac{\gamma^{\alpha} \left[\esl_{2T} \psl_2 \phi^T_{\rho}(x_{2})\right] \gamma_{\mu} \psl_1
\gamma^{\nu} (\psl_1-\ksl_2 + \lsl + m_b) \gamma_{\alpha} (\psl_1-\ksl_1+\lsl+m_b) \gamma_{\nu}  (\psl_1)
\gamma_{5}\left[\frac{\nsl_+}{\sqrt{2}}\phi^+_B(x_1) \right] }
{[(p_1-k_2)^2-m^2_b](k_1-k_2)^2[(p_1-k_2+l)^2-m^2_b][(p_1-k_1+l)^2-m^2_b] l^2}  \bigg \} \non
&=& -\frac{1}{8} \phi^{(1)}_{B,d} \otimes \left[G^{(0)}_{bT,1}(x_1,x_2)-G^{(0)}_{bT,1}(\xi_1,x_1,x_2) \right],    \\
	%%%%%%
G^{(1)}_{bT1,2g} &=& -\frac{eg^4}{9}
\mathbf{Tr} \bigg \{ \frac{\gamma^{\nu} (\ksl_1 - \lsl) \gamma^{\alpha} \left[\esl_{2T} \psl_2 \phi^T_{\rho}(x_{2})\right]
\gamma_{\mu} \psl_1 \gamma_{\nu} (\psl_1-\ksl_2-\lsl+m_b)  \gamma_{\alpha}  (\psl_1)\gamma_{5}\left[\frac{\nsl_+}{\sqrt{2}}\phi^+_B(x_1) \right] }
{[(p_1-k_2)^2-m^2_b](k_1-k_2-l)^2[(p_1-k_2-l)^2-m^2_b](k_1-l)^2 l^2}  \bigg \} \non
&&\sim 0,  \\
	%%%%%%
	G^{(1)}_{bT1,2h} &=& \frac{eg^4}{9}
	\mathbf{Tr} \bigg \{  \frac{\gamma^{\alpha} \left[\esl_{2T} \psl_2 \phi^T_{\rho}(x_{2})\right] \gamma^{\nu}  (\psl_2-\ksl_2+\lsl)
\gamma_{\mu} (\psl_1 +\lsl ) \gamma_{\alpha}  (\psl_1-\ksl_1+\lsl+m_b) \gamma_{\nu} (\psl_1) \gamma_{5}\left[\frac{\nsl_+}{\sqrt{2}}\phi^+_B(x_1) \right] }
	{[(p_1-k_2+l)^2-m^2_b](k_1-k_2)^2(p_2-k_2+l)^2[(p_1-k_1+l)^2-m^2_b] l^2}  \bigg \} \non
	&&\sim 0,  \\
	%%%%%%
	G^{(1)}_{bT1,2i} &=& -\frac{eg^4}{9}
\mathbf{Tr} \bigg \{  \frac{\gamma^{\alpha} (\ksl_2+\lsl) \gamma^{\nu} \left[\esl_{2T} \psl_2 \phi^T_{\rho}(x_{2})\right]
\gamma_{\mu} \psl_1 \gamma_{\alpha} (\psl_1-\ksl_1+\lsl+m_b) \gamma_{\nu} (\psl_1) \gamma_{5}\left[\frac{\nsl_+}{\sqrt{2}}\phi^+_B(x_1) \right] }
	{[(p_1-k_2)^2-m^2_b](k_1-k_2+l)^2(k_2+l)^2[(p_1-k_1+l)^2-m^2_b] l^2}  \bigg \} \non
	&&\sim 0,  \\
	%%%%%%
	G^{(1)}_{bT1,2j} &=&  \frac{eg^4}{9}
	\mathbf{Tr} \bigg \{  \frac{\gamma^{\nu} (\ksl_1-\lsl) \gamma^{\alpha} (\ksl_2-\lsl) \gamma_{\nu} \left[\esl_{2T}
\psl_2 \phi^T_{\rho}(x_{2})\right] \gamma_{\mu} \psl_1 \gamma_{\alpha} (\psl_1) \gamma_{5}\left[\frac{\nsl_+}{\sqrt{2}}\phi^+_B(x_1) \right] }
	{[(p_1-k_2)^2-m^2_b](k_1-k_2)^2(k_1-l)^2(k_2-l)^2 l^2}  \bigg \} \non
	&=&   \frac{1}{8} \phi^{(1)}_{B,e} \otimes G^{(0)}_{bT,1}(x_1,x_2),  \\
	%%%%%%
	G^{(1)}_{bT1,2k} &=& \frac{eg^4}{9}
	\mathbf{Tr} \bigg \{  \frac{\gamma^{\nu} (\ksl_1-\lsl) \gamma^{\alpha} \left[\esl_{2T} \psl_2 \phi^T_{\rho}(x_{2})\right]
\gamma_{\nu}(\psl_2-\ksl_2-\lsl) \gamma_{\mu} (\psl_1-\lsl) \gamma_{\alpha} (\psl_1) \gamma_{5}\left[\frac{\nsl_+}{\sqrt{2}}\phi^+_B(x_1)\right]}
	{[(p_1-k_2-l)^2-m^2_b](k_1-k_2-l)^2(p_2-k_2-l)^2(k_1-l)^2l^2}  \bigg \} \non
	&=&   -\frac{1}{8} \phi^{(1)}_{B,e} \otimes G^{(0)}_{bT,1}(\xi_1,x_2),	\label{eq:GbT1f-k}
\eeq}
with the wave functions $\phi^{(1)}_{B,d}$  and $\phi^{(1)}_{B,e}$ are of the following form:
{\small
\beq
\phi^{(1)}_{B,d} &=& \frac{-ig^2C_F}{8} \frac{ \gamma^+\gamma_{5}\gamma^-  (\psl_1-\ksl_1+\lsl+m_b) \gamma^{\rho}
\gamma^+\gamma_{5}\gamma^- }{[(p_1-k_1+l)^2-m^2_b] l^2} \frac{\nu_{\rho}}{\nu \cdot l},  \\
	%%%%
\phi^{(1)}_{B,e} &=& \frac{ig^2C_F}{8} \frac{ \gamma^+\gamma_{5}\gamma^- \gamma^{\rho} (\ksl_1-\lsl)  \gamma^+
\gamma_{5}\gamma^- }{(k_1-l)^2 l^2} \frac{\nu_{\rho}}{\nu \cdot l}.  \label{eq:WFbT1de}
\eeq}

The direct summation of above contributions  gives the following result:
{\small
	\beq
	G^{(1)}_{up,bT3}(x_1,x_2) &=& \phi^{(1)}_{B,d}(x_1,\xi_1) \otimes \left[G^{(0)}_{bT}(x_1,x_2)-G^{(0)}_{bT}(x_1,\xi_1,x_2) \right] \non
	&&+ \frac{1}{8}\phi^{(1)}_{B,d}(x_1,\xi_1) \otimes  \left[G^{(0)}_{bT}(\xi_1,x_1,x_2)-G^{(0)}_{bT}(x_1,\xi_1,x_2) \right],
	\label{eq:bT1up} \\
	G^{(1)}_{down,aL}(x_1,x_2) &=& \phi^{(1)}_{B,e}(x_1,\xi_1) \otimes
	\left[G^{(0)}_{aL}(x_1,x_2)-G^{(0)}_{aL}(x_1,\xi_1,x_2) \right],
	\label{eq:bT1down}
	\eeq}
where the $1/8$ term seems abnormal here. Fortunately, as we demonstrated in Eqs.~(\ref{eq:GaLxi-1}-\ref{eq:GaLxi-3}) ,
the difference between
$G^{(0)}_{bT}(\xi_1,x_1,x_2)$ and $G^{(0)}_{bT}(x_1,\xi_1,x_2)$ is the  form of the propagator in denominator.
The flow $l$ here can be recognized as a momentum shift for
$x'_1=x_1- l^+/p_1$ and with the symmetry by choice
of $P_2$ and $k_2$, the infrared structure between the two terms will be canceled each other.

%%%%%%%%%%%%%%%%%%%%%%%%%%%%%%%%%%%%%%%%%%%%%%%%%%%%%%%%%%%%%%%%%%%%%%%%%%%%
%%%%%%%%%%%%%%%%%%%%%%%%%%%%%%%%%%%%%%%%%%%%%%%%%%%%%%%%%%%%%%%%%%%%%%%%%%%%

\subsection{The NLO amplitudes for $G^{(0)}_{bT2}$ }

The amplitudes for the NLO contributions from the irreducible diagrams to  $G^{(0)}_{bT2}$ are the following:
{\small
	\beq
	G^{(1)}_{bT2,2d} &=& -eg^4
	\mathbf{Tr} \bigg{\{} \frac{\gamma^{\alpha} \left[\esl_{2T} \psl_2 \phi^T_{\rho}(x_{2})\right] \gamma_{\mu} \psl_1  \gamma^{\beta} (\psl_1-\ksl_1+\lsl+m_b)
 \gamma^{\gamma} (\psl_1) \gamma_{5}\left[\frac{\nsl_-}{\sqrt{2}}\phi^-_B(x_1)\right] F_{\alpha \beta \gamma} }
	{(k_1-k_2)^2(k_1-k_2-l)^2[(p_1-k_2)^2-m^2_b][(p_1-k_1+l)^2-m^2_b]l^2}  \bigg{\}}   \non
	&=& \frac{9}{8} \phi^{(1)}_{B,d} \otimes \left[G^{(0)}_{bT,2}(x_1,x_2)-G^{(0)}_{bT,2}(x_1,\xi_1,x_2) \right], \\
	%%%%%
	G^{(1)}_{bT2,2e} &=& eg^4
	\mathbf{Tr} \bigg{\{} \frac{\gamma^{\alpha}(\ksl_1-\lsl) \gamma^{\beta} \left[\esl_{2T} \psl_2 \phi^T_{\rho}(x_{2})\right]
\gamma_{\mu} \psl_1 \gamma^{\gamma}  (\psl_1)\gamma_{5}\left[\frac{\nsl_-}{\sqrt{2}}\phi^-_B(x_1)\right] F_{\alpha \beta \gamma} }
	{(k_1-k_2)^2(k_1-k_2-l)^2[(p_1-k_2)^2-m^2_b](k_1-l)^2l^2}  \bigg{\}}   \non
	&&\sim 0 ,	\label{eq:GbT2de}
	\eeq}
{\small
	\beq
	G^{(1)}_{bT2,2f} &=& \frac{eg^4}{9}
    \mathbf{Tr} \bigg{\{} \frac{\gamma^{\alpha} \left[\esl_{2T} \psl_2 \phi^T_{\rho}(x_{2})\right] \gamma_{\mu} \psl_1
    \gamma^{\nu} (\psl_1-\ksl_2 + \lsl + m_b) \gamma_{\alpha} (\psl_1-\ksl_1+\lsl+m_b) \gamma_{\nu}  (\psl_1)
    \gamma_{5}\left[\frac{\nsl_-}{\sqrt{2}}\phi^-_B(x_1)\right] }
	{[(p_1-k_2)^2-m^2_b](k_1-k_2)^2[(p_1-k_2+l)^2-m^2_b][(p_1-k_1+l)^2-m^2_b] l^2}  \bigg{\}} \non
	&=& -\frac{1}{8} \phi^{(1)}_{B,d} \otimes \left[G^{(0)}_{bT,2}(x_1,x_2)-G^{(0)}_{bT,2}(\xi_1,x_1,x_2) \right] ,  \\
	%%%%%%
	G^{(1)}_{bT2,2g} &=& -\frac{eg^4}{9}
	\mathbf{Tr} \bigg{\{} \frac{\gamma^{\nu} (\ksl_1 - \lsl) \gamma^{\alpha} \left[\esl_{2T} \psl_2 \phi^T_{\rho}(x_{2})\right]
\gamma_{\mu} \psl_1 \gamma_{\nu} (\psl_1-\ksl_2-\lsl+m_b)  \gamma_{\alpha}  (\psl_1)\gamma_{5}\left[\frac{\nsl_-}{\sqrt{2}}\phi^-_B(x_1)\right] }
	{[(p_1-k_2)^2-m^2_b](k_1-k_2-l)^2[(p_1-k_2-l)^2-m^2_b](k_1-l)^2 l^2}  \bigg{\}} \non
	&&\sim 0, \\
	%%%%%%
	G^{(1)}_{bT2,2h} &=&  \frac{eg^4}{9}
	\mathbf{Tr} \bigg{\{} \frac{\gamma^{\alpha} \left[\esl_{2T} \psl_2 \phi^T_{\rho}(x_{2})\right] \gamma^{\nu}  (\psl_2-\ksl_2+\lsl)
\gamma_{\mu} (\psl_1 +\lsl ) \gamma_{\alpha}  (\psl_1-\ksl_1+\lsl+m_b) \gamma_{\nu} (\psl_1) \gamma_{5}\left[\frac{\nsl_-}{\sqrt{2}}\phi^-_B(x_1)\right] }
	{[(p_1-k_2+l)^2-m^2_b](k_1-k_2)^2(p_2-k_2+l)^2[(p_1-k_1+l)^2-m^2_b] l^2}  \bigg{\}} \non
	&=&   -\frac{1}{8} \phi^{(1)}_{B,d} \otimes G^{(0)}_{bT,2}(\xi_1,x_1,x_2) ,\\
	%%%%%%
	G^{(1)}_{bT2,2i} &=&  -\frac{eg^4}{9}
	\mathbf{Tr} \bigg{\{} \frac{\gamma^{\alpha} (\ksl_2+\lsl) \gamma^{\nu} \left[\esl_{2T} \psl_2 \phi^T_{\rho}(x_{2})\right] \gamma_{\mu} \psl_1
\gamma_{\alpha} (\psl_1-\ksl_1+\lsl+m_b) \gamma_{\nu} (\psl_1) \gamma_{5}\left[\frac{\nsl_-}{\sqrt{2}}\phi^-_B(x_1)\right] }
	{[(p_1-k_2)^2-m^2_b](k_1-k_2+l)^2(k_2+l)^2[(p_1-k_1+l)^2-m^2_b] l^2}  \bigg{\}} \non
	&=&   \frac{1}{8} \phi^{(1)}_{B,d} \otimes G^{(0)}_{bT,2}(x_1,\xi_1,x_2), \\
	%%%%%%
	G^{(1)}_{bT2,2j} &=& -\frac{eg^4}{9}
	\mathbf{Tr} \bigg{\{} \frac{\gamma^{\nu} (\ksl_1-\lsl) \gamma^{\alpha} (\ksl_2-\lsl) \gamma_{\nu} \left[\esl_{2T} \psl_2 \phi^T_{\rho}(x_{2})\right]
\gamma_{\mu} \psl_1 \gamma_{\alpha} (\psl_1) \gamma_{5}\left[\frac{\nsl_-}{\sqrt{2}}\phi^-_B(x_1)\right] }
	{[(p_1-k_2)^2-m^2_b](k_1-k_2)^2(k_1-l)^2(k_2-l)^2 l^2}  \bigg{\}} \non
	&&\sim 0 ,\\
	%%%%%%
	G^{(1)}_{bT2,2k} &=&  \frac{eg^4}{9}
	\mathbf{Tr} \bigg{\{} \frac{\gamma^{\nu} (\ksl_1-\lsl) \gamma^{\alpha} \left[\esl_{2T} \psl_2 \phi^T_{\rho}(x_{2})\right]  \gamma_{\nu}(\psl_2-\ksl_2-\lsl)
\gamma_{\mu} (\psl_1-\lsl) \gamma_{\alpha} (\psl_1) \gamma_{5}\left[\frac{\nsl_-}{\sqrt{2}}\phi^-_B(x_1)\right]}
	{[(p_1-k_2-l)^2-m^2_b](k_1-k_2-l)^2(p_2-k_2-l)^2(k_1-l)^2l^2}  \bigg{\}} \non
	&&\sim 0, 	\label{eq:GbT2f-k}
	\eeq}
with the wave function $	\phi^{(1)}_{B,d} $ as the form of
{\small
\beq
	\phi^{(1)}_{B,d} &=& \frac{-ig^2C_F}{8} \frac{ \gamma^-\gamma_{5}\gamma^+  (\psl_1-\ksl_1+\lsl+m_b) \gamma^{\rho}
\gamma^-\gamma_{5}\gamma^+ }{[(p_1-k_1+l)^2-m^2_b] l^2} \frac{\nu_{\rho}}{\nu \cdot l}.
	\label{eq:WFbT2d}
	\eeq}
The summation of the contributions from these irreducible sub-diagrams gives the following result:
{\small
	\beq
G^{(1)}_{up,bT2}(x_1,x_2) = \phi^{(1)}_{B,d}(x_1,\xi_1) \otimes \left[G^{(0)}_{bT}(x_1,x_2)-G^{(0)}_{bT}(x_1,\xi_1,x_2) \right].
	\label{eq:bT2up}
\eeq}

%%%%%%%%%%%%%%%%%%%%%%%%%%%%%%%%%%%%%%%%%%%%%%%%%%%%%%%%%%%%%%%%%%%%%%%%%%%%%%%
%%%%%%%%%%%%%%%%%%%%%%%%%%%%%%%%%%%%%%%%%%%%%%%%%%%%%%%%%%%%%%%%%%%%%%%%%%%%%%%
\subsection{The NLO amplitudes for $G^{(0)}_{bT3}$ }

The amplitudes for the NLO contributions from the irreducible diagrams to  $G^{(0)}_{bT3}$ are the following:
{\small
	\beq
	G^{(1)}_{bT3,2d} &=& -eg^4
	\mathbf{Tr} \bigg{\{} \frac{\gamma^{\alpha}\left[\esl_{2T} \psl_2 \phi^T_{\rho}(x_{2})\right] \gamma_{\mu} m_b  \gamma^{\beta} (\psl_1-\ksl_1+\lsl+m_b)
\gamma^{\gamma} (m_B) \gamma_{5}\left[\frac{\nsl_+}{\sqrt{2}}\phi^+_B(x_1)\right] F_{\alpha \beta \gamma} }
	{(k_1-k_2)^2(k_1-k_2-l)^2[(p_1-k_2)^2-m^2_b][(p_1-k_1+l)^2-m^2_b]l^2}  \bigg{\}}   \non
	&=& \frac{9}{8} \phi^{(1)}_{B,d} \otimes \left[G^{(0)}_{bT,3}(x_1,x_2)-G^{(0)}_{bT,3}(x_1,\xi_1,x_2) \right], \\
	%%%%%
	G^{(1)}_{bT3,2e} &=& eg^4
	\mathbf{Tr} \bigg{\{} \frac{\gamma^{\alpha}(\ksl_1-\lsl) \gamma^{\beta} \left[\esl_{2T} \psl_2 \phi^T_{\rho}(x_{2})\right] \gamma_{\mu} m_b
\gamma^{\gamma}  (m_B)\gamma_{5}\left[\frac{\nsl_+}{\sqrt{2}}\phi^+_B(x_1)\right] F_{\alpha \beta \gamma} }
	{(k_1-k_2)^2(k_1-k_2-l)^2[(p_1-k_2)^2-m^2_b](k_1-l)^2l^2}  \bigg{\}}   \non
	&=& \frac{9}{8} \phi^{(1)}_{B,e} \otimes \left[G^{(0)}_{bT,3}(x_1,x_2)-G^{(0)}_{bT,3}(x_1,\xi_1,x_2) \right],
	\label{eq:GbT3de}
	\eeq}
{\small
	\beq
	G^{(1)}_{bT3,2f} &=& \frac{eg^4}{9}
	\mathbf{Tr} \bigg{\{} \frac{\gamma^{\alpha} \left[\esl_{2T} \psl_2 \phi^T_{\rho}(x_{2})\right] \gamma_{\mu} m_b  \gamma^{\nu} (\psl_1-\ksl_2 + \lsl + m_b)
\gamma_{\alpha} (\psl_1-\ksl_1+\lsl+m_b) \gamma_{\nu}  (m_B) \gamma_{5}\left[\frac{\nsl_+}{\sqrt{2}}\phi^+_B(x_1) \right] }
	{[(p_1-k_2)^2-m^2_b](k_1-k_2)^2[(p_1-k_2+l)^2-m^2_b][(p_1-k_1+l)^2-m^2_b] l^2}  \bigg{\}} \non
	&=& -\frac{1}{8} \phi^{(1)}_{B,d} \otimes \left[G^{(0)}_{bT,3}(x_1,x_2)-G^{(0)}_{bT,3}(\xi_1,x_1,x_2) \right],    \\
	%%%%%%
	G^{(1)}_{bT3,2g} &=& -\frac{eg^4}{9}
	\mathbf{Tr} \bigg{\{} \frac{\gamma^{\nu} (\ksl_1 - \lsl) \gamma^{\alpha} \left[\esl_{2T} \psl_2 \phi^T_{\rho}(x_{2})\right] \gamma_{\mu} m_b
\gamma_{\nu} (\psl_1-\ksl_2-\lsl+m_b)  \gamma_{\alpha}  (m_B)\gamma_{5}\left[\frac{\nsl_+}{\sqrt{2}}\phi^+_B(x_1) \right] }
	{[(p_1-k_2)^2-m^2_b](k_1-k_2-l)^2[(p_1-k_2-l)^2-m^2_b](k_1-l)^2 l^2}  \bigg{\}} \non
	&=& -\frac{1}{8} \phi^{(1)}_{B,e}  \otimes \left[G^{(0)}_{bT,3}(x_1,\xi_1,x_2)-G^{(0)}_{bT,3}(\xi_1,x_2) \right],  \\
	%%%%%%
	G^{(1)}_{bT3,2h} &=&  \frac{eg^4}{9}
	\mathbf{Tr} \bigg{\{} \frac{\gamma^{\alpha} \left[\esl_{2T} \psl_2 \phi^T_{\rho}(x_{2})\right] \gamma^{\nu}  (\psl_2-\ksl_2+\lsl)
\gamma_{\mu} (m_b +\lsl ) \gamma_{\alpha}  (\psl_1-\ksl_1+\lsl+m_b) \gamma_{\nu} (m_B) \gamma_{5}\left[\frac{\nsl_+}{\sqrt{2}}\phi^+_B(x_1) \right] }
	{[(p_1-k_2+l)^2-m^2_b](k_1-k_2)^2(p_2-k_2+l)^2[(p_1-k_1+l)^2-m^2_b] l^2}  \bigg{\}} \non
	&=&   -\frac{1}{8} \phi^{(1)}_{B,d} \otimes G^{(0)}_{bT,3}(\xi_1,x_1,x_2) , \\
	%%%%%%
	G^{(1)}_{bT3,2i} &=&  -\frac{eg^4}{9}
	\mathbf{Tr} \bigg{\{} \frac{\gamma^{\alpha} (\ksl_2+\lsl) \gamma^{\nu} \left[\esl_{2T} \psl_2 \phi^T_{\rho}(x_{2})\right]
\gamma_{\mu} m_b \gamma_{\alpha} (\psl_1-\ksl_1+\lsl+m_b) \gamma_{\nu} (m_B) \gamma_{5}\left[\frac{\nsl_+}{\sqrt{2}}\phi^+_B(x_1) \right] }
	{[(p_1-k_2)^2-m^2_b](k_1-k_2+l)^2(k_2+l)^2[(p_1-k_1+l)^2-m^2_b] l^2}  \bigg{\}} \non
	&=&   \frac{1}{8} \phi^{(1)}_{B,d} \otimes G^{(0)}_{bT,3}(x_1,\xi_1,x_2),\\
	%%%%%%
	G^{(1)}_{bT3,2j} &=&  \frac{eg^4}{9}
	\mathbf{Tr} \bigg{\{} \frac{\gamma^{\nu} (\ksl_1-\lsl) \gamma^{\alpha} (\ksl_2-\lsl) \gamma_{\nu} \left[\esl_{2T} \psl_2 \phi^T_{\rho}(x_{2})\right]
\gamma_{\mu} m_b \gamma_{\alpha} (m_B) \gamma_{5}\left[\frac{\nsl_+}{\sqrt{2}}\phi^+_B(x_1) \right] }
	{[(p_1-k_2)^2-m^2_b](k_1-k_2)^2(k_1-l)^2(k_2-l)^2 l^2}  \bigg{\}} \non
	&=&   \frac{1}{8} \phi^{(1)}_{B,e} \otimes G^{(0)}_{bT,3}(x_1,x_2), \\
	%%%%%%
	G^{(1)}_{bT3,2k} &=&  -\frac{eg^4}{9}
	\mathbf{Tr} \bigg{\{} \frac{\gamma^{\nu} (\ksl_1-\lsl) \gamma^{\alpha} \left[\esl_{2T} \psl_2 \phi^T_{\rho}(x_{2})\right]  \gamma_{\nu}(\psl_2-\ksl_2-\lsl)
\gamma_{\mu} (m_b-\lsl) \gamma_{\alpha} (m_B) \gamma_{5}\left[\frac{\nsl_+}{\sqrt{2}}\phi^+_B(x_1)\right]}
	{[(p_1-k_2-l)^2-m^2_b](k_1-k_2-l)^2(p_2-k_2-l)^2(k_1-l)^2l^2}  \bigg{\}} \non
	&=&   -\frac{1}{8} \phi^{(1)}_{B,e} \otimes G^{(0)}_{bT,3}(\xi_1,x_2),
	\label{eq:GbT3f-k}
	\eeq}
with the wave functions $ \phi^{(1)}_{B,d}$ and $ \phi^{(1)}_{B,e}$ in the form of
{\small
	\beq
	\phi^{(1)}_{B,d} &=& \frac{-ig^2C_F}{4} \frac{ \gamma^+\gamma_{5}  (\psl_1-\ksl_1+\lsl+m_b) \gamma^{\rho} \gamma_{5}
\gamma^- }{[(p_1-k_1+l)^2-m^2_b] l^2} \frac{\nu_{\rho}}{\nu \cdot l} , \non
	%%%%
	\phi^{(1)}_{B,e} &=& \frac{ig^2C_F}{4} \frac{ \gamma_{5}\gamma^- \gamma^{\rho} (\ksl_1-\lsl)  \gamma^+
\gamma_{5} }{(k_1-l)^2 l^2} \frac{\nu_{\rho}}{\nu \cdot l} . 	\label{eq:WFbT3de}
	\eeq}
The summation of above terms from the irreducible sub-diagrams gives the following result:
{\small
	\beq
	G^{(1)}_{up,bT3}(x_1,x_2) = \phi^{(1)}_{B,d}(x_1,\xi_1) \otimes \left[G^{(0)}_{bT}(x_1,x_2)-G^{(0)}_{bT}(x_1,\xi_1,x_2) \right]
	\label{eq:bT3up}, \\
	G^{(1)}_{down,aL}(x_1,x_2) = \phi^{(1)}_{B,e}(x_1,\xi_1) \otimes
	\left[G^{(0)}_{aL}(x_1,x_2)-G^{(0)}_{aL}(x_1,\xi_1,x_2) \right].
	\label{eq:bT3down}
	\eeq}

%%%%%%%%%%%%%%%%%%%%%%%%%%%%%%%%%%%%%%%%%%%%%%%%%%%%%%%%%%%%%%%%%%%%%%%%%%%%%%%

\subsection{The NLO amplitudes for $G^{(0')}_{bL2}$ }

The amplitudes for the NLO contributions from the irreducible diagrams to  $G^{(0')}_{bL2}$ with the additional gluon
emitted from the final $\rho$ meson are the following:
{\small
\beq
G^{(1')}_{bL2,2d} &=& -eg^4
	\mathbf{Tr} \bigg{\{} \frac{\gamma^{\alpha}\left[M_{\rho} \esl_{2L} \phi_{\rho}(x_{2})\right] \gamma^{\beta} (\psl_2-\ksl_2-\lsl) \gamma_{\mu} m_b
\gamma^{\gamma} (\psl_1) \gamma_{5}\left[\frac{\nsl_-}{\sqrt{2}}\phi^-_B(x_1)\right] F_{\alpha \beta \gamma} }
	{(k_1-k_2)^2(k_1-k_2-l)^2[(p_1-k_2-l)^2-m^2_b](p_2-k_2-l)^2 l^2}  \bigg{\}}   \non
	&=& \frac{9}{8} \phi^{(1)}_{\rho,d} \otimes \left[G^{(0)}_{bL,2}(x_1,\xi_2,x_2)-G^{(0)}_{bL,2}(x_1,\xi_2) \right], \\
	%%%%%
	G^{(1')}_{bL2,2e} &=& eg^4
	\mathbf{Tr} \bigg{\{} \frac{\gamma^{\alpha}(\ksl_2+\lsl) \gamma^{\beta} \left[M_{\rho} \esl_{2L} \phi_{\rho}(x_{2})\right]  \gamma_{\mu} m_b
\gamma^{\gamma}  (\psl_1) \gamma_{5}\left[\frac{\nsl_-}{\sqrt{2}}\phi^-_B(x_1)\right] F_{\alpha \beta \gamma} }
	{(k_1-k_2)^2(k_1-k_2-l)^2[(p_1-k_2)^2-m^2_b](k_2+l)^2l^2}  \bigg{\}}   \non
	&=& \frac{9}{8} \phi^{(1)}_{\rho,e} \otimes \left[-G^{(0)}_{bL,2}(x_1,x_2)+G^{(0)}_{bL,2}(x_1,x_2,\xi_2) \right],
	\label{eq:GbL2'de}
	\eeq}
{\small
	\beq
	G^{(1')}_{bL2,2f} &=& -\frac{eg^4 C^2_F}{2}
	\mathbf{Tr} \bigg{\{} \frac{\gamma^{\alpha} \left[M_{\rho} \esl_{2L} \phi_{\rho}(x_{2})\right]  \gamma^{\nu} (\psl_2-\ksl_2-\lsl)
\gamma_{\mu}  (\psl_1-\ksl_2 -\lsl + m_b) \gamma_{\nu}  m_b \gamma_{\alpha}  (\psl_1) \gamma_{5}\left[\frac{\nsl_-}{\sqrt{2}}\phi^-_B(x_1)\right] }
	{[(p_1-k_2)^2-m^2_b](k_1-k_2)^2[(p_1-k_2-l)^2-m^2_b](p_2-k_2-l)^2 l^2}  \bigg{\}} \non
	&&\sim 0,   \\
	%%%%%%
	G^{(1')}_{bL2,2g} &=& \frac{eg^4 C^2_F}{2}
	\mathbf{Tr} \bigg{\{} \frac{ \gamma^{\alpha}(\ksl_2 + \lsl) \gamma^{\nu} \left[M_{\rho} \esl_{2L} \phi_{\rho}(x_{2})\right]  \gamma_{\mu} m_b
\gamma_{\nu} (\psl_1-\ksl_2-\lsl+m_b)  \gamma_{\alpha}  (\psl_1) \gamma_{5}\left[\frac{\nsl_-}{\sqrt{2}}\phi^-_B(x_1)\right] }
	{[(p_1-k_2)^2-m^2_b](k_1-k_2-l)^2[(p_1-k_2-l)^2-m^2_b](k_2+l)^2 l^2}  \bigg{\}} \non
	&=& \phi^{(1)}_{\rho ,e}  \otimes \left[-G^{(0)}_{bL,2}(x_1,x_2,\xi_2)+G^{(0)}_{bL,2}(x_1,\xi_2) \right],
\eeq }
{\small
\beq
	%%%%%%
	G^{(1')}_{bL2,2h} &=&  \frac{eg^4}{9}
	\mathbf{Tr} \bigg{\{} \frac{\gamma^{\alpha} \left[M_{\rho} \esl_{2L} \phi_{\rho}(x_{2})\right]  \gamma^{\nu}  (\psl_2-\ksl_2 -\lsl) \gamma_{\mu} (m_b -\lsl )
\gamma_{\alpha}  (\psl_1-\ksl_1-\lsl+m_b) \gamma_{\nu} (\psl_1) \gamma_{5}\left[\frac{\nsl_-}{\sqrt{2}}\phi^-_B(x_1)\right] }
	{[(p_1-k_2-l)^2-m^2_b](k_1-k_2)^2(p_2-k_2-l)^2[(p_1-k_1-l)^2-m^2_b] l^2}  \bigg{\}} \non
	&=&   -\frac{1}{8} \phi^{(1)}_{\rho,d} \otimes G^{(0)}_{bL,2}(x_1,\xi_2,x_2),  \\
	%%%%%%
	G^{(1')}_{bL2,2i} &=&  -\frac{eg^4}{9}
	\mathbf{Tr} \bigg{\{} \frac{\gamma^{\nu} (\ksl_1-\lsl) \gamma^{\alpha}  \left[M_{\rho} \esl_{2L} \phi_{\rho}(x_{2})\right] \gamma_{\nu} (\psl_2-\ksl_2-\lsl)
\gamma_{\mu} (m_b-\lsl) \gamma_{\alpha}   (\psl_1) \gamma_{5}\left[\frac{\nsl_-}{\sqrt{2}}\phi^-_B(x_1)\right] }
	{[(p_1-k_2-l)^2-m^2_b](k_1-k_2-l)^2(p_2-k_2-l)^2 (k_1-l)^2 l^2}  \bigg{\}} \non
	&&\sim 0,
\eeq }
{\small
\beq
	%%%%%%
	G^{(1')}_{bL2,2j} &=&  \frac{eg^4}{9}
	\mathbf{Tr} \bigg{\{} \frac{\gamma^{\nu} (\ksl_1+\lsl) \gamma^{\alpha} (\ksl_2+\lsl) \gamma_{\nu} \left[M_{\rho} \esl_{2L} \phi_{\rho}(x_{2})\right]
 \gamma_{\mu} m_b \gamma_{\alpha} (\psl_1) \gamma_{5}\left[\frac{\nsl_-}{\sqrt{2}}\phi^-_B(x_1)\right] }
	{[(p_1-k_2)^2-m^2_b](k_1-k_2)^2(k_1+l)^2(k_2+l)^2 l^2}  \bigg{\}} \non
    &&\sim 0, \\
	%%%%%%
	G^{(1')}_{bL2,2k} &=&  -\frac{eg^4}{9}
	\mathbf{Tr} \bigg{\{} \frac{\gamma^{\alpha} (\ksl_2+\lsl) \gamma^{\nu} \left[M_{\rho} \esl_{2L} \phi_{\rho}(x_{2})\right] \gamma_{\mu} m_b
\gamma_{\alpha} (\psl_1-\ksl_1+\lsl+m_b) \gamma_{\nu} (\psl_1) \gamma_{5}\left[\frac{\nsl_-}{\sqrt{2}}\phi^-_B(x_1)\right]}
	{[(p_1-k_2)^2-m^2_b](k_1-k_2-l)^2[(p_1-k_1+l)^2-m^2_b](k_2+l)^2l^2}  \bigg{\}} \non
	&=&   -\frac{1}{8} \phi^{(1)}_{\rho,e} \otimes G^{(0)}_{bL,2}(x_1,x_2,\xi_2) ,
	\label{eq:GbL2'f-k}
	\eeq}
with the wave functions $ \phi^{(1)}_{\rho,d}$ and $ \phi^{(1)}_{\rho,e}$ in the form of
{\small
\beq
\phi^{(1)}_{\rho,d} &=& \frac{-ig^2C_F}{4} \frac{ \gamma^+ \gamma^{\rho} (\psl_2-\ksl_2-\lsl)  \gamma^- }{(p_2-k_2-l)^2 l^2}
\frac{n_{\rho}}{n \cdot l}, \non
	%%%%
\phi^{(1)}_{\rho,e} &=& \frac{ig^2C_F}{4} \frac{ \gamma^-  (\ksl_2+\lsl) \gamma^{\rho}  \gamma^+ }{(k_2+l)^2 l^2}
\frac{n_{\rho}}{n \cdot l}. \label{eq:WFbL2'de}
	\eeq}
Analogous to the ones in Eq.~(\ref{eq:bT1up}),  the $1/8$ terms also appear in this channel.
After the summation of the contributions from these irreducible sub-diagrams, one finds the following result:
{\small
	\beq
	G^{(1')}_{up,bL2}(x_1,x_2) = \phi^{(1)}_{\rho,d} \otimes \left[G^{(0)}_{bL,2}(x_1,\xi_2,x_2)-G^{(0)}_{bL,2}(x_1,\xi_2) \right], \\
	G^{(1')}_{down,bL2}(x_1,x_2) = \phi^{(1)}_{\rho,e} \otimes \left[-G^{(0)}_{bL,2}(x_1,x_2)+G^{(0)}_{bL,2}(x_1,x_2,\xi_2) \right].
	\label{eq:bL1'down}
	\eeq}

%%%%%%%%%%%%%%%%%%%%%%%%%%%%%%%%%%%%%%%%%%%%%%%%%%%%%%%%%%%%%%%%%%%%%%%%%%%%%%%
%%%%%%%%%%%%%%%%%%%%%%%%%%%%%%%%%%%%%%%%%%%%%%%%%%%%%%%%%%%%%%%%%%%%%%%%%%%%%%%
\subsection{The NLO amplitudes for $G^{(0')}_{bT1}$ }

The amplitudes for the NLO contributions from the irreducible diagrams to  $G^{(0')}_{bT1}$ with the additional gluon
emitted from the final $\rho$ meson are the following:
{\small
	\beq
	G^{(1')}_{bT1,2d} &=& -eg^4
	\mathbf{Tr} \bigg{\{} \frac{\gamma^{\alpha} \left[\esl_{2T} \psl_2 \phi^T_{\rho}(x_{2})\right]  \gamma^{\beta} (\psl_2-\ksl_2-\lsl)
\gamma_{\mu} \psl_1  \gamma^{\gamma} (\psl_1) \gamma_{5}\left[\frac{\nsl_+}{\sqrt{2}}\phi^+_B(x_1)\right] F_{\alpha \beta \gamma} }
	{(k_1-k_2)^2(k_1-k_2-l)^2[(p_1-k_2-l)^2-m^2_b](p_2-k_2-l)^2 l^2}  \bigg{\}}   \non
	&=& \frac{9}{8} \phi^{(1)}_{\rho,d} \otimes \left[G^{(0)}_{bT,1}(x_1,\xi_2,x_2)-G^{(0)}_{bT,1}(x_1,\xi_2) \right], \\
	%%%%%
	G^{(1')}_{bT1,2e} &=& eg^4
	\mathbf{Tr} \bigg{\{} \frac{\gamma^{\alpha}(\ksl_2+\lsl) \gamma^{\beta} \left[\esl_{2T} \psl_2 \phi^T_{\rho}(x_{2})\right]   \gamma_{\mu} \psl_1
\gamma^{\gamma}  (\psl_1) \gamma_{5}\left[\frac{\nsl_+}{\sqrt{2}}\phi^+_B(x_1)\right] F_{\alpha \beta \gamma} }
	{(k_1-k_2)^2(k_1-k_2-l)^2[(p_1-k_2)^2-m^2_b](k_2+l)^2l^2}  \bigg{\}}   \non
	&=& \frac{9}{8} \phi^{(1)}_{\rho,e} \otimes \left[-G^{(0)}_{bT,1}(x_1,x_2)+G^{(0)}_{bT,1}(x_1,x_2,\xi_2) \right], \label{eq:GbT1'de}
	\eeq}
{\small
	\beq
	G^{(1')}_{bT1,2f} &=& -\frac{eg^4 C^2_F}{2}
	\mathbf{Tr} \bigg{\{} \frac{\gamma^{\alpha} \left[\esl_{2T} \psl_2 \phi^T_{\rho}(x_{2})\right]  \gamma^{\nu} (\psl_2-\ksl_2-\lsl)
\gamma_{\mu}  (\psl_1-\ksl_2 -\lsl + m_b) \gamma_{\nu}  \psl_1 \gamma_{\alpha}  (\psl_1) \gamma_{5}\left[\frac{\nsl_+}{\sqrt{2}}\phi^+_B(x_1)\right] }
	{[(p_1-k_2)^2-m^2_b](k_1-k_2)^2[(p_1-k_2-l)^2-m^2_b](p_2-k_2-l)^2 l^2}  \bigg{\}} \non
	&=& \phi^{(1)}_{\rho,d} \otimes \left[G^{(0)}_{bT,1}(x_1,x_2)-G^{(0)}_{bT,1}(x_1,\xi_2,x_2) \right], \\
	%%%%%%
	G^{(1')}_{bT1,2g} &=& \frac{eg^4 C^2_F}{2}
	\mathbf{Tr} \bigg{\{} \frac{ \gamma^{\alpha}(\ksl_2 + \lsl) \gamma^{\nu} \left[\esl_{2T} \psl_2 \phi^T_{\rho}(x_{2})\right]
\gamma_{\mu} \psl_1 \gamma_{\nu} (\psl_1-\ksl_2-\lsl+m_b)  \gamma_{\alpha}  (\psl_1) \gamma_{5}\left[\frac{\nsl_+}{\sqrt{2}}\phi^+_B(x_1)\right] }
	{[(p_1-k_2)^2-m^2_b](k_1-k_2-l)^2[(p_1-k_2-l)^2-m^2_b](k_2+l)^2 l^2}  \bigg{\}} \non
	&=&  \phi^{(1)}_{\rho ,e}  \otimes \left[-G^{(0)}_{bT,1}(x_1,x_2,\xi_2)+G^{(0)}_{bT,1}(x_1,\xi_2) \right] ,
\eeq }
	%%%%%%
{\small
\beq
	G^{(1')}_{bT1,2h} &=&  \frac{eg^4}{9}
	\mathbf{Tr} \bigg{\{} \frac{\gamma^{\alpha} \left[\esl_{2T} \psl_2 \phi^T_{\rho}(x_{2})\right]  \gamma^{\nu}  (\psl_2-\ksl_2+\lsl)
\gamma_{\mu} (\psl_1 +\lsl ) \gamma_{\alpha}  (\psl_1-\ksl_1+\lsl+m_b) \gamma_{\nu} (\psl_1) \gamma_{5}\left[\frac{\nsl_+}{\sqrt{2}}\phi^+_B(x_1)\right] }
	{[(p_1-k_2+l)^2-m^2_b](k_1-k_2)^2(p_2-k_2+l)^2[(p_1-k_1+l)^2-m^2_b] l^2}  \bigg{\}} \non
	&&\sim 0 , \\
	%%%%%%
	G^{(1')}_{bT1,2i} &=&  -\frac{eg^4}{9}
	\mathbf{Tr} \bigg{\{} \frac{\gamma^{\nu} (\ksl_1-\lsl) \gamma^{\alpha}  \left[\esl_{2T} \psl_2 \phi^T_{\rho}(x_{2})\right]
\gamma_{\nu} (\psl_2-\ksl_2-\lsl)\gamma_{\mu} (\psl_1-\lsl) \gamma_{\alpha}   (\psl_1) \gamma_{5}\left[\frac{\nsl_+}{\sqrt{2}}\phi^+_B(x_1)\right] }
	{[(p_1-k_2-l)^2-m^2_b](k_1-k_2-l)^2(p_2-k_2-l)^2 (k_1-l)^2 l^2}  \bigg{\}} \non
	&=& \frac{1}{8} \phi^{(1)}_{\rho ,d}  \otimes G^{(0)}_{bT,1}(x_1,\xi_2),
\eeq}
	%%%%%%
{\small
\beq
	G^{(1')}_{bT1,2j} &=&  \frac{eg^4}{9}
	\mathbf{Tr} \bigg{\{} \frac{\gamma^{\nu} (\ksl_1+\lsl) \gamma^{\alpha} (\ksl_2+\lsl) \gamma_{\nu} \left[\esl_{2T} \psl_2 \phi^T_{\rho}(x_{2})\right]
 \gamma_{\mu} \psl_1 \gamma_{\alpha} (\psl_1) \gamma_{5}\left[\frac{\nsl_+}{\sqrt{2}}\phi^+_B(x_1)\right]}
	{[(p_1-k_2)^2-m^2_b](k_1-k_2)^2(k_1+l)^2(k_2+l)^2 l^2}  \bigg{\}} \non
	&=& \frac{1}{8} \phi^{(1)}_{\rho ,e}  \otimes G^{(0)}_{bT,2}(x_1,x_2),  \\
	%%%%%%
	G^{(1')}_{bT1,2k} &=&  -\frac{eg^4}{9}
	\mathbf{Tr} \bigg{\{} \frac{\gamma^{\alpha} (\ksl_2+\lsl) \gamma^{\nu} \left[\esl_{2T} \psl_2 \phi^T_{\rho}(x_{2})\right]  \gamma_{\mu} \psl_1
\gamma_{\alpha} (\psl_1-\ksl_1+\lsl+m_b) \gamma_{\nu} (\psl_1) \gamma_{5}\left[\frac{\nsl_+}{\sqrt{2}}\phi^+_B(x_1)\right]}
	{[(p_1-k_2)^2-m^2_b](k_1-k_2-l)^2[(p_1-k_1+l)^2-m^2_b](k_2+l)^2l^2}  \bigg{\}} \non
	&&\sim 0, 	\label{eq:GbT1'f-k}
	\eeq}
with the wave functions $ \phi^{(1)}_{\rho,d}$ and $ \phi^{(1)}_{\rho,e}$ in the form of
{\small
	\beq
	\phi^{(1)}_{\rho,d} &=& \frac{-ig^2C_F}{4} \frac{\gamma_{\perp} \gamma^+  \gamma^{\rho} (\psl_2-\ksl_2-\lsl) \gamma^-
\gamma_{\perp} }{(p_2-k_2-l)^2 l^2} \frac{n_{\rho}}{n \cdot l} \non
	%%%%
	\phi^{(1)}_{\rho,e} &=& \frac{ig^2C_F}{4} \frac{ \gamma^-\gamma_{\perp}  (\ksl_2+\lsl) \gamma^{\rho}  \gamma_{\perp}
\gamma^+ }{(k_2+l)^2 l^2} \frac{n_{\rho}}{n \cdot l}.
	\label{eq:WFbT1'de}
	\eeq}
The  summation of   the contributions from above irreducible sub-diagrams gives the following result:
{\small
	\beq
	G^{(1')}_{up,bT1}(x_1,x_2) = \phi^{(1)}_{\rho,d} \otimes \left[G^{(0)}_{bT,1}(x_1,\xi_2,x_2)-G^{(0)}_{bT,1}(x_1,\xi_2) \right],
	\label{eq:bT1'up} \\
	G^{(1')}_{down,bT1}(x_1,x_2) = \phi^{(1)}_{\rho,e} \otimes \left[-G^{(0)}_{bT,1}(x_1,x_2)+G^{(0)}_{bT,1}(x_1,x_2,\xi_2) \right].
	\label{eq:bT1'down}
	\eeq}

%%%%%%%%%%%%%%%%%%%%%%%%%%%%%%%%%%%%%%%%%%%%%%%%%%%%%%%%%%%%%%%%%%%%%%%%%%%%%%%
%%%%%%%%%%%%%%%%%%%%%%%%%%%%%%%%%%%%%%%%%%%%%%%%%%%%%%%%%%%%%%%%%%%%%%%%%%%%%%%
\subsection{The NLO amplitudes for $G^{(0')}_{bT2}$ }  %% 7

The NLO amplitudes for $G^{(0')}_{bT2}$ with the additional gluon emitted from the
final state $\rho$ meson are the following:
{\small
	\beq
	G^{(1')}_{bT2,2d} &=& -eg^4
	\mathbf{Tr} \bigg{\{} \frac{\gamma^{\alpha} \left[\esl_{2T} \psl_2 \phi^T_{\rho}(x_{2})\right]  \gamma^{\beta} (\psl_2-\ksl_2-\lsl)
 \gamma_{\mu} \psl_1  \gamma^{\gamma} (\psl_1) \gamma_{5}\left[\frac{\nsl_-}{\sqrt{2}}\phi^-_B(x_1)\right] F_{\alpha \beta \gamma} }
	{(k_1-k_2)^2(k_1-k_2-l)^2[(p_1-k_2-l)^2-m^2_b](p_2-k_2-l)^2 l^2}  \bigg{\}}   \non
	&=& \frac{9}{8} \phi^{(1)}_{\rho,d} \otimes \left[G^{(0)}_{bT,2}(x_1,\xi_2,x_2)-G^{(0)}_{bT,2}(x_1,\xi_2) \right] , \\
	%%%%%
	G^{(1')}_{bT2,2e} &=& eg^4
	\mathbf{Tr} \bigg{\{} \frac{\gamma^{\alpha}(\ksl_2+\lsl) \gamma^{\beta} \left[\esl_{2T} \psl_2 \phi^T_{\rho}(x_{2})\right]
\gamma_{\mu} \psl_1 \gamma^{\gamma}  (\psl_1) \gamma_{5} \left[\frac{\nsl_-}{\sqrt{2}}\phi^-_B(x_1)\right] F_{\alpha \beta \gamma} }
	{(k_1-k_2)^2(k_1-k_2-l)^2[(p_1-k_2)^2-m^2_b](k_2+l)^2l^2}  \bigg{\}}   \non
	&=& \frac{9}{8} \phi^{(1)}_{\rho,e} \otimes \left[-G^{(0)}_{bT,2}(x_1,x_2)+G^{(0)}_{bT,2}(x_1,x_2,\xi_2) \right],
	\label{eq:GbT2'de}
	\eeq}
{\small
	\beq
	G^{(1')}_{bT2,2f} &=& -\frac{eg^4 C^2_F}{2}
	\mathbf{Tr} \bigg{\{} \frac{\gamma^{\alpha} \left[\esl_{2T} \psl_2 \phi^T_{\rho}(x_{2})\right]  \gamma^{\nu} (\psl_2-\ksl_2-\lsl)
\gamma_{\mu}  (\psl_1-\ksl_2 -\lsl + m_b) \gamma_{\nu}  \psl_1 \gamma_{\alpha}  (\psl_1) \gamma_{5}\left[\frac{\nsl_-}{\sqrt{2}}\phi^-_B(x_1)\right] }
	{[(p_1-k_2)^2-m^2_b](k_1-k_2)^2[(p_1-k_2-l)^2-m^2_b](p_2-k_2-l)^2 l^2}  \bigg{\}} \non
	&=& \phi^{(1)}_{\rho,d} \otimes \left[G^{(0)}_{bT,2}(x_1,x_2)-G^{(0)}_{bT,2}(x_1,\xi_2,x_2) \right] ,  \\
	%%%%%%
	G^{(1')}_{bT2,2g} &=& \frac{eg^4 C^2_F}{2}
	\mathbf{Tr} \bigg{\{} \frac{ \gamma^{\alpha}(\ksl_2 + \lsl) \gamma^{\nu} \left[\esl_{2T} \psl_2 \phi^T_{\rho}(x_{2})\right]   \gamma_{\mu} \psl_1
\gamma_{\nu} (\psl_1-\ksl_2-\lsl+m_b)  \gamma_{\alpha}  (\psl_1) \gamma_{5}\left[\frac{\nsl_-}{\sqrt{2}}\phi^-_B(x_1)\right] }
	{[(p_1-k_2)^2-m^2_b](k_1-k_2-l)^2[(p_1-k_2-l)^2-m^2_b](k_2+l)^2 l^2}  \bigg{\}} \non
	&=&  \phi^{(1)}_{\rho ,e}  \otimes \left[-G^{(0)}_{bT,2}(x_1,x_2,\xi_2)+G^{(0)}_{bT,2}(x_1,\xi_2) \right] ,
\eeq}
	%%%%%%
{\small
\beq	
G^{(1')}_{bT2,2h} &=&  \frac{eg^4}{9}
	\mathbf{Tr} \bigg{\{} \frac{\gamma^{\alpha} \left[\esl_{2T} \psl_2 \phi^T_{\rho}(x_{2})\right]  \gamma^{\nu}  (\psl_2-\ksl_2-\lsl)
\gamma_{\mu} (\psl_1 +\lsl ) \gamma_{\alpha}  (\psl_1-\ksl_1-\lsl+m_b) \gamma_{\nu} (\psl_1) \gamma_{5}\left[\frac{\nsl_-}{\sqrt{2}}\phi^-_B(x_1)\right] }
	{[(p_1-k_2+l)^2-m^2_b](k_1-k_2)^2(p_2-k_2+l)^2[(p_1-k_1-l)^2-m^2_b] l^2}  \bigg{\}} \non
	&=& -\frac{1}{8}\phi^{(1)}_{\rho,d} \otimes G^{(0)}_{bT,2}(x_1,\xi_2,x_2),  \\
	%%%%%%
	G^{(1')}_{bT2,2i} &=&  -\frac{eg^4}{9}
	\mathbf{Tr} \bigg{\{} \frac{\gamma^{\nu} (\ksl_1-\lsl) \gamma^{\alpha}  \left[\esl_{2T} \psl_2 \phi^T_{\rho}(x_{2})\right]
\gamma_{\nu} (\psl_2-\ksl_2-\lsl)\gamma_{\mu} (\psl_1-\lsl) \gamma_{\alpha}   (\psl_1) \gamma_{5}\left[\frac{\nsl_-}{\sqrt{2}}\phi^-_B(x_1)\right] }
	{[(p_1-k_2-l)^2-m^2_b](k_1-k_2-l)^2(p_2-k_2-l)^2 (k_1-l)^2 l^2}  \bigg{\}} \non
	&& \sim 0,
\eeq}
{\small
\beq
	%%%%%%
	G^{(1')}_{bT2,2j} &=&  \frac{eg^4}{9}
	\mathbf{Tr} \bigg{\{} \frac{\gamma^{\nu} (\ksl_1+\lsl) \gamma^{\alpha} (\ksl_2+\lsl) \gamma_{\nu} \left[\esl_{2T} \psl_2 \phi^T_{\rho}(x_{2})\right]
\gamma_{\mu} \psl_1 \gamma_{\alpha} (\psl_1) \gamma_{5}\left[\frac{\nsl_-}{\sqrt{2}}\phi^-_B(x_1)\right]}
	{[(p_1-k_2)^2-m^2_b](k_1-k_2)^2(k_1+l)^2(k_2+l)^2 l^2}  \bigg{\}} \non
	&& \sim 0, \\
	%%%%%%
	G^{(1')}_{bT2,2k} &=&  -\frac{eg^4}{9}
	\mathbf{Tr} \bigg{\{} \frac{\gamma^{\alpha} (\ksl_2+\lsl) \gamma^{\nu} \left[\esl_{2T} \psl_2 \phi^T_{\rho}(x_{2})\right]  \gamma_{\mu} \psl_1
\gamma_{\alpha} (\psl_1-\ksl_1+\lsl+m_b) \gamma_{\nu} (\psl_1) \gamma_{5}\left[\frac{\nsl_-}{\sqrt{2}}\phi^-_B(x_1)\right]}
	{[(p_1-k_2)^2-m^2_b](k_1-k_2-l)^2[(p_1-k_1+l)^2-m^2_b](k_2+l)^2l^2}  \bigg{\}} \non
	&=&  - \frac{1}{8}\phi^{(1)}_{\rho ,e}  \otimes  G^{(0)}_{bT,2}(x_1,x_2,\xi_2),
	\label{eq:GbT2'f-k}
	\eeq}
with the wave functions $ \phi^{(1)}_{\rho,d}$ and $ \phi^{(1)}_{\rho,e}$ in the form of
{\small
	\beq
	\phi^{(1)}_{\rho,d} &=& \frac{-ig^2C_F}{4} \frac{\gamma_{\perp} \gamma^+  \gamma^{\rho} (\psl_2-\ksl_2-\lsl) \gamma^-
\gamma_{\perp} }{(p_2-k_2-l)^2 l^2} \frac{n_{\rho}}{n \cdot l},  \non
	%%%%
	\phi^{(1)}_{\rho,e} &=& \frac{ig^2C_F}{4} \frac{ \gamma^-\gamma_{\perp}  (\ksl_2+\lsl) \gamma^{\rho}  \gamma_{\perp}
\gamma^+ }{(k_2+l)^2 l^2} \frac{n_{\rho}}{n \cdot l} .
	\label{eq:WFbT2'de}
	\eeq}
The summation over  the contributions from above irreducible sub-diagrams gives the following result:
{\small
	\beq
	G^{(1')}_{up,bT2}(x_1,x_2) = \phi^{(1)}_{\rho,d} \otimes \left[G^{(0)}_{bT,2}(x_1,\xi_2,x_2)-G^{(0)}_{bT,2}(x_1,\xi_2) \right],
	\label{eq:bT2'up} \\
	G^{(1')}_{down,bT1}(x_1,x_2) = \phi^{(1)}_{\rho,e} \otimes \left[-G^{(0)}_{bT,2}(x_1,x_2)+G^{(0)}_{bT,2}(x_1,x_2,\xi_2) \right].
	\label{eq:bT2'down}
	\eeq}

%%%%%%%%%%%%%%%%%%%%%%%%%%%%%%%%%%%%%%%%%%%%%%%%%%%%%%%%%%%%%%%%%%%%%%%%%%%%%%%
%%%%%%%%%%%%%%%%%%%%%%%%%%%%%%%%%%%%%%%%%%%%%%%%%%%%%%%%%%%%%%%%%%%%%%%%%%%%%%%

\subsection{The NLO amplitudes for $G^{(0')}_{bT3}$ }   %% 8

The NLO amplitudes for $G^{(0')}_{bT3}$ with the additional gluon emitted from the final $\rho$ meson are the following:
{\small
	\beq
	G^{(1')}_{bT3,2d} &=& -eg^4
	\mathbf{Tr} \bigg{\{} \frac{\gamma^{\alpha} \left[\esl_{2T} \psl_2 \phi^T_{\rho}(x_{2})\right]  \gamma^{\beta} (\psl_2-\ksl_2-\lsl) \gamma_{\mu} m_b
\gamma^{\gamma} (m_B) \gamma_{5}\left[\frac{\nsl_+}{\sqrt{2}}\phi^+_B(x_1)\right] F_{\alpha \beta \gamma} }
	{(k_1-k_2)^2(k_1-k_2-l)^2[(p_1-k_2-l)^2-m^2_b](p_2-k_2-l)^2 l^2}  \bigg{\}}   \non
	&=& \frac{9}{8} \phi^{(1)}_{\rho,d} \otimes \left[G^{(0)}_{bT,3}(x_1,\xi_2,x_2)-G^{(0)}_{bT,3}(x_1,\xi_2) \right], \\
	%%%%%
	G^{(1')}_{bT3,2e} &=& eg^4
	\mathbf{Tr} \bigg{\{} \frac{\gamma^{\alpha}(\ksl_2+\lsl) \gamma^{\beta} \left[\esl_{2T} \psl_2 \phi^T_{\rho}(x_{2})\right]   \gamma_{\mu} m_b
\gamma^{\gamma}  (m_B) \gamma_{5}\left[\frac{\nsl_+}{\sqrt{2}}\phi^+_B(x_1)\right] F_{\alpha \beta \gamma} }
	{(k_1-k_2)^2(k_1-k_2-l)^2[(p_1-k_2)^2-m^2_b](k_2+l)^2l^2}  \bigg{\}}   \non
	&=& \frac{9}{8} \phi^{(1)}_{\rho,e} \otimes \left[-G^{(0)}_{bT,3}(x_1,x_2)+G^{(0)}_{bT,3}(x_1,x_2,\xi_2) \right],
	\label{eq:GbT3'de}
	\eeq}
{\small
	\beq
	G^{(1')}_{bT3,2f} &=& -\frac{eg^4 C^2_F}{2}
	\mathbf{Tr} \bigg{\{} \frac{\gamma^{\alpha} \left[\esl_{2T} \psl_2 \phi^T_{\rho}(x_{2})\right]  \gamma^{\nu} (\psl_2-\ksl_2-\lsl)
\gamma_{\mu}  (\psl_1-\ksl_2 -\lsl + m_b) \gamma_{\nu}  m_b \gamma_{\alpha}  (m_B) \gamma_{5}\left[\frac{\nsl_+}{\sqrt{2}}\phi^+_B(x_1)\right] }
	{[(p_1-k_2)^2-m^2_b](k_1-k_2)^2[(p_1-k_2-l)^2-m^2_b](p_2-k_2-l)^2 l^2}  \bigg{\}} \non
	&=& \phi^{(1)}_{\rho,d} \otimes \left[G^{(0)}_{bT,3}(x_1,x_2)-G^{(0)}_{bT,3}(x_1,\xi_2,x_2) \right] ,  \\
	%%%%%%
	G^{(1')}_{bT3,2g} &=& \frac{eg^4 C^2_F}{2}
	\mathbf{Tr} \bigg{\{} \frac{ \gamma^{\alpha}(\ksl_2 + \lsl) \gamma^{\nu} \left[\esl_{2T} \psl_2 \phi^T_{\rho}(x_{2})\right]
\gamma_{\mu} m_b \gamma_{\nu} (\psl_1-\ksl_2-\lsl+m_b)  \gamma_{\alpha}  (m_B) \gamma_{5}\left[\frac{\nsl_+}{\sqrt{2}}\phi^+_B(x_1)\right] }
	{[(p_1-k_2)^2-m^2_b](k_1-k_2-l)^2[(p_1-k_2-l)^2-m^2_b](k_2+l)^2 l^2}  \bigg{\}} \non
	&& \sim 0 ,
\eeq}
{\small
\beq
	G^{(1')}_{bT3,2h} &=&  \frac{eg^4}{9}
	\mathbf{Tr} \bigg{\{} \frac{\gamma^{\alpha} \left[\esl_{2T} \psl_2 \phi^T_{\rho}(x_{2})\right]  \gamma^{\nu}  (\psl_2-\ksl_2-\lsl)
\gamma_{\mu} (m_b -\lsl ) \gamma_{\alpha}  (\psl_1-\ksl_1-\lsl+m_b) \gamma_{\nu} (m_B) \gamma_{5}\left[\frac{\nsl_+}{\sqrt{2}}\phi^+_B(x_1)\right] }
	{[(p_1-k_2-l)^2-m^2_b](k_1-k_2)^2(p_2-k_2-l)^2[(p_1-k_1-l)^2-m^2_b] l^2}  \bigg{\}} \non
	&=& -\phi^{(1)}_{\rho,d} \otimes G^{(0)}_{bT,3}(x_1,\xi_2,x_2), \\
	%%%%%%
	G^{(1')}_{bT3,2i} &=&  -\frac{eg^4}{9}
	\mathbf{Tr} \bigg{\{} \frac{\gamma^{\nu} (\ksl_1-\lsl) \gamma^{\alpha}  \left[\esl_{2T} \psl_2 \phi^T_{\rho}(x_{2})\right]
\gamma_{\nu} (\psl_2-\ksl_2-\lsl)\gamma_{\mu} (m_b-\lsl) \gamma_{\alpha}   (m_B) \gamma_{5}\left[\frac{\nsl_+}{\sqrt{2}}\phi^+_B(x_1)\right] }
	{[(p_1-k_2-l)^2-m^2_b](k_1-k_2-l)^2(p_2-k_2-l)^2 (k_1-l)^2 l^2}  \bigg{\}} \non
	&=& \frac{1}{8} \phi^{(1)}_{\rho ,d}  \otimes G^{(0)}_{bT,3}(x_1,\xi_2),
\eeq}
{\small
\beq
	G^{(1')}_{bT3,2j} &=&  \frac{eg^4}{9}
	\mathbf{Tr} \bigg{\{} \frac{\gamma^{\nu} (\ksl_1+\lsl) \gamma^{\alpha} (\ksl_2+\lsl) \gamma_{\nu} \left[\esl_{2T} \psl_2 \phi^T_{\rho}(x_{2})\right]
\gamma_{\mu} m_b \gamma_{\alpha} (m_B) \gamma_{5}\left[\frac{\nsl_+}{\sqrt{2}}\phi^+_B(x_1)\right]}
	{[(p_1-k_2)^2-m^2_b](k_1-k_2)^2(k_1+l)^2(k_2+l)^2 l^2}  \bigg{\}} \non
	&=& \frac{1}{8} \phi^{(1)}_{\rho ,e}  \otimes G^{(0)}_{bT,3}(x_1,x_2),  \\
	%%%%%%
	G^{(1')}_{bT3,2k} &=&  -\frac{eg^4}{9}
	\mathbf{Tr} \bigg{\{} \frac{\gamma^{\alpha} (\ksl_2+\lsl) \gamma^{\nu} \left[\esl_{2T} \psl_2 \phi^T_{\rho}(x_{2})\right]
\gamma_{\mu} m_b \gamma_{\alpha} (\psl_1-\ksl_1+\lsl+m_b) \gamma_{\nu} (m_B) \gamma_{5}\left[\frac{\nsl_+}{\sqrt{2}}\phi^+_B(x_1)\right]}
	{[(p_1-k_2)^2-m^2_b](k_1-k_2-l)^2[(p_1-k_1+l)^2-m^2_b](k_2+l)^2l^2}  \bigg{\}} \non
	&=& -\frac{1}{8} \phi^{(1)}_{\rho ,e}  \otimes G^{(0)}_{bT,3}(x_1,x_2,\xi_2) , 	\label{eq:GbT3'f-k}
	\eeq}
with the wave functions $ \phi^{(1)}_{\rho,d}$ and $ \phi^{(1)}_{\rho,e}$ in the form of
{\small
	\beq
	\phi^{(1)}_{\rho,d} &=& \frac{-ig^2C_F}{4} \frac{\gamma_{\perp} \gamma^+  \gamma^{\rho} (\psl_2-\ksl_2-\lsl) \gamma^-
\gamma_{\perp} }{(p_2-k_2-l)^2 l^2} \frac{n_{\rho}}{n \cdot l},  \non
	%%%%
	\phi^{(1)}_{\rho,e} &=& \frac{ig^2C_F}{4} \frac{ \gamma^-\gamma_{\perp}  (\ksl_2+\lsl) \gamma^{\rho}  \gamma_{\perp}
\gamma^+ }{(k_2+l)^2 l^2} \frac{n_{\rho}}{n \cdot l}.
	\label{eq:WFbT3'de}
	\eeq}
The summation of the contributions from above irreducible sub-diagrams gives the following result:
{\small
	\beq
	G^{(1')}_{up,bT3}(x_1,x_2) = \phi^{(1)}_{\rho,d} \otimes \left[G^{(0)}_{bT,3}(x_1,\xi_2,x_2)-G^{(0)}_{bT,3}(x_1,\xi_2) \right],
	\label{eq:bT3'up} \\
	G^{(1')}_{down,bT3}(x_1,x_2) = \phi^{(1)}_{\rho,e} \otimes \left[-G^{(0)}_{bT,3}(x_1,x_2)+G^{(0)}_{bT,3}(x_1,x_2,\xi_2) \right].
	\label{eq:bT3'down}
	\eeq}

\end{appendix}
%%%%%%%%%%%%%%%%%%%%%%%%%%%%%%%%%%%%%%%%%%%%%%%%%%%%%%%%%%%%%%%%%%%%%%%%%%%%%%%
%%%%%%%%%%%%%%%%%%%%%%%%%%%%%%%%%%%%%%%%%%%%%%%%%%%%%%%%%%%%%%%%%%%%%%%%%%%%%%%

\end{document}